\documentclass[amsmath,endfloats,preprint,prb,11pt,nobibnotes]{revtex4}

\usepackage{graphicx}
\usepackage{bm}

\usepackage{hyperref}
\hypersetup{breaklinks=true}

\begin{document}

\title{
       An exact equilibrium reduced density matrix formulation I:
       The influence of noise, disorder, and temperature on 
       localization in excitonic systems
       }

\author{
       Jeremy M. Moix
       } 
\affiliation{
       School of Materials Science and Engineering,
       Nanyang Technological University
       Singapore 639798
       }
\affiliation{
       Department of Chemistry,
       Massachusetts Institute of Technology,
       77 Massachusetts Avenue, Cambridge, MA 02139
       }
\author{
       Yang Zhao
       } 
\affiliation{
       School of Materials Science and Engineering,
       Nanyang Technological University
       Singapore 639798
       }
\author{
       Jianshu Cao
       } \email{jianshu@mit.edu}
\affiliation{
       Department of Chemistry,
       Massachusetts Institute of Technology,
       77 Massachusetts Avenue, Cambridge, MA 02139
       }

\date{\today}

\begin{abstract}

An exact method to compute the entire equilibrium reduced density matrix 
for systems characterized by a system-bath Hamiltonian is presented.
The approach is based upon a stochastic unraveling of the influence functional 
that appears in the imaginary time path integral formalism of quantum 
statistical mechanics.
This method is then applied to study the effects of thermal noise, 
static disorder, and temperature on the coherence length in excitonic systems.
As representative examples of biased and unbiased systems,
attention is focused on the well-characterized light harvesting complexes 
of FMO and LH2, respectively.
Due to the bias, FMO  is completely localized in the site basis 
at low temperatures, whereas LH2 is completely delocalized.
In the latter, the presence of static disorder leads to a plateau in the 
coherence length at low temperature that becomes increasingly pronounced
with increasing strength of the disorder.
The introduction of noise, however, precludes this effect.
In biased systems, it is shown that the environment may increase the 
coherence length, but only decrease that of unbiased systems.
Finally it is emphasized that for typical values of the environmental 
parameters in light harvesting systems, the system and bath are 
entangled at equilibrium in the single excitation manifold.
That is, the density matrix cannot be described as a product
state as is often assumed, even at room temperature.
The reduced density matrix of LH2 is shown to be in precise agreement with 
the steady state limit of previous exact quantum dynamics calculations.

\end{abstract}

\maketitle

\section{Introduction}

Only very few systems may be considered as truly isolated.
Typically a source of dephasing and dissipation is present due
to the interaction of the system with its surrounding environment.
In this case, the most relevant physical quantity is the reduced density
operator, obtained by tracing the total Boltzmann operator over the 
environmental degrees of freedom. 
Because of the interaction with the environment the reduced density matrix 
is, in general, not equal to that obtained from the Boltzmann distribution of 
the system alone.
However, despite its broad importance, very few numerically exact methods 
are available to provide the entire reduced density matrix in a simple and
efficient manner.
Perhaps the most common approach is based upon the imaginary time version
of the path integral formulation of quantum mechanics.\cite{kleinert04,schul86}
Imaginary time path integral methods can treat truly macroscopic
environments through the influence functional techniques introduced by 
Feynman and Vernon.\cite{feynman63,Ingold88} 
However, due to the nature of the boundary value problem, 
an independent calculation must be performed for every element of the density
matrix which quickly becomes prohibitive for large systems.
As a result most path integral calculations focus specifically
on the partition function, from which many equilibrium properties 
may be obtained, rather than the entire density matrix.

Due to the increasing ability of experimental techniques to probe
aspects of small quantum systems and the coherences within them, 
knowledge of the partition function alone is becoming insufficient 
for many applications. 
For example, the off-diagonal elements of the reduced density matrix
determine the amount of entanglement in quantum information 
and quantum computing applications, as well as the amount of quantum 
coherence present in photosynthetic systems.
While only indirectly physically realizable, these quantities are 
of increasing practical importance.
For instance, it has been shown that the energy transport in photosynthetic
systems is most efficient in a regime that lies between the limiting cases of
fully quantum transport and completely classical hopping.\cite{wu10,wu11}
Additionally, the reduced density matrix serves as the initial state 
for numerical simulations of the dynamics of open quantum systems.
Oftentimes this initial state is assumed to factorize into a product of 
independent system and bath states.
It is well known that this approximation is not always valid, and the error
introduced increases both as the temperature is lowered and as the 
system-bath coupling increases.
One of the main focuses of this work is a general path integral formulation
for the entire reduced density matrix in open quantum systems.

After the exposition of the formalism, we then systematically 
assess the effects of the environment on the reduced density matrix,
focusing on two model photosynthetic light harvesting systems.
In particular, the Fenna-Matthews-Olson (FMO) protein and the 
light harvesting complex (LH2) of photosynthetic bacteria, 
both of which play a key role in the energy transfer process.
These complexes consist of several individual chromophores that are 
closely spaced and strongly coupled, which allows for very high quantum 
efficiencies.
While composed of identical chromophores, LH2 and FMO play fundamentally 
different roles in the photosynthetic process.
FMO serves as an energy funnel that transfers excitonic energy from a light
harvesting antenna on to the reaction center where charge separation occurs.
As a result its energy landscape is highly biased to facilitate this process.
LH2, on the other hand, is a highly symmetric antenna complex whose
function is to gather solar energy.
In contrast to FMO, it possesses a nearly homogeneous energy profile.
While both complexes are essential for photosynthesis, here they serve as 
examples of biased and unbiased excitonic systems with reasonably 
well-characterized Hamiltonians, and for which a wealth of experimental 
and numerical results are available.

In such complicated biological systems, the role of the environment 
must be included.
Each light harvesting complex is subject to both static disorder arising 
from different local environments (inhomogeneous broadening), 
as well as thermal noise originating from coupling of the exciton 
to the phonon bath (homogeneous broadening). 
Estimates of the energy scales involved in the exciton transfer place light
harvesting systems in an interesting regime.\cite{cho05}
The coupling between nearest neighbors in the system is of the same
order of magnitude as the exciton-phonon coupling, which in turn, is 
comparable to the thermal energy as well as to the strength of static disorder.
That is, all the relevant interactions must be properly taken into account
even to develop a qualitative description of these systems.

In photosynthetic systems, the strong coupling between chromophores 
leads to excitons that are delocalized over several of the individual sites.
The physical extent of the exciton is referred to as the coherence length and 
is encoded in the off-diagonal elements of the reduced density matrix 
in the site basis.
The coherence length plays an important role in determining
the spectroscopic properties of light harvesting systems.
The most pronounced signature of which is related to superradiance, 
a collective phenomenon where the radiative rate of an
aggregate is enhanced with respect to that of a single 
monomer.\cite{meier97,meier97a,monshouwer97,potma98}
Superradiance has been observed in LH2, resulting in approximately a
three-fold increase of the radiative rate and lending further support 
for the Frenkel exciton description of light harvesting 
systems.\cite{monshouwer97}
In addition to the superradiance, the coherence length has also been
linked to the Stokes shift at low temperatures.\cite{chernyak99}

While the coherence length has no precise definition,
there have been several theoretical and experimental attempts to 
quantify this quantity in excitonic systems in order to assess the 
relative importance of the experimentally observed quantum coherences in 
the energy transfer 
process.\cite{pullerits96,kuhn97,ray99,strumpfer09,yakolev02,cheng06,jang01}
Perhaps not surprisingly, the estimates vary widely depending on the 
techniques used and the particular definition employed. 
In LH2 for example, these estimates range from a lower limit of 
2-3 chromophores to extending over almost the entire 18 sites of the system.
In numerical calculations, the effects of temperature and static disorder 
are relatively straightforward to include.
An extensive comparison of many different measures for the coherence length 
has been presented for LH2 in the presence of static disorder.\cite{dahlbom01}
Others have used direct comparisons of the wavefunctions or the 
density matrices in order to develop a qualitative understanding of how disorder
leads to localization.\cite{prokhorenko03, mulken07, meier97, yakolev02}
The effects of the particular choice of the distribution of static disorder 
on the coherence length as well as the role of diagonal and 
off-diagonal disorder in one-dimensional chains and in extended tubular 
aggregate systems have also been 
explored.\cite{fidder91, dijkstra08, eisfeld10, vlaming09}
However, the influence of thermal noise on the coherence length
is much more difficult to treat, and is generally either completely neglected 
or included only approximately.\cite{kumble98,meier97,meier97a,novoderezhkin98}
Notable exceptions include the exact calculations of Ray and Makri 
in which they used the size of the bead from 
imaginary time path integral calculations as yet another estimate 
for the coherence length in LH2.\cite{ray99}
Similarly, Ishizaki and Fleming have used the exact hierarchy 
equations of motion simulations to study the concurrence and its
time-dependence in a two-site model for LHCII.\cite{ishizaki10a}
Here we present exact results treating both the thermal bath and
the effects of static disorder.

There are two central focuses of this work.
In Sec.~\ref{sec:Methods} an exact method is presented 
to efficiently calculate the entire equilibrium reduced density matrix for
open systems characterized by a system-bath Hamiltonian.
While this scheme is designed with excitonic systems in mind,
it is applicable to a much broader class of systems.
The approach is based on a stochastic unraveling of the influence functional 
in the imaginary time path integral.\cite{chandler91, schul86, cao96, cao96a}
This leads to independent calculations of the density matrix driven by 
stochastic colored noise.
Averaging over the noise distribution then provides the exact reduced 
density matrix of the open system.
Because of the stochastic nature of the algorithm, the process readily lends 
itself to a straightforward and efficient Monte Carlo procedure.
The approach is valid for arbitrary spectral densities of the bath as 
well as the simultaneous presence of static disorder.
This provides a practical route to investigate the role of the bath
in rather large systems.

Using this method, the influence of the environment on the coherence 
length in the LH2 and FMO is then systematically investigated in 
Sec.~\ref{sec:Results}.
For two different commonly used definitions of the coherence length,
numerically exact results are presented for LH2 and FMO.
It is shown that these two systems display qualitatively different 
behavior at low temperatures. 
At low temperatures and in the absence absent of noise or disorder,
symmetric quantum systems, such as LH2, are fully coherent and delocalized over 
the entire domain.
The effect of the environment in this case is to localize the exciton 
over a small subset of the chromophores.
However, for biased systems such as FMO, localization always occurs at
sufficiently low temperature even in the absence of noise or disorder.
In this case the Boltzmann weighting becomes the dominant factor 
localizing the population on the lowest energy site.
Here the presence of noise and disorder can increase the coherence length.

For typical values of the system-bath coupling, 
it is demonstrated that the system and the bath are entangled at equilibrium
in the single exciton manifold.
This effect modifies the equilibrium populations and becomes more important
as the system-bath coupling increases and as the temperature is lowered.
The equilibrium distribution can not be written as 
a separable product of system and bath states as is often assumed.
In unbiased systems, static disorder leads to a plateau in the coherence 
length at low temperatures, which becomes more pronounced as the strength 
of the disorder is increased.
Introducing a thermal bath, however, delays the onset of such a feature
and may prevent its formation altogether.
This is consistent with recent experimental measurements.\cite{kamishima07}
Additionally, we make an important clarification regarding the origin 
of static disorder.
Often in light harvesting systems, it is stated that static disorder
arises from the low frequency dynamics of the protein environment, 
i.e., a slow bath.
It is shown here that true static disorder leads to a steady decrease of the 
coherence length at low temperature whereas a slow bath causes a drastic 
change in the localization length in this regime.

\section{Methods} \label{sec:Methods}

The total Hamiltonian is decomposed into a sum of system, bath, 
and system-bath interactions,
\begin{equation}
   \hat H = \hat H_s + \hat H_b + \hat H_{sb}
   \;.\label{eq:Ham}
\end{equation}
In this section we are primarily concerned with the treatment of the bath
so for clarity, we will consider only the example 
of a dissipative, one-dimensional continuous system.
However, the generalization to multi-dimensional systems is 
straightforward and will become readily apparent below.
Details for discrete systems, 
which are the focus of the numerical results below, 
are discussed in Appendix~\ref{sec:Discrete_PI} and follow accordingly.
For this case, the system Hamiltonian is given by,
\begin{equation}
   \hat H_s = \frac{1}{2M} \dot {\hat q}^2 + V(\hat q) \;,
\end{equation}
where $M$ is the mass of the particle and $V$ is an arbitrary potential.
The bath consists of an infinite set of harmonic oscillators 
which are linearly coupled to the system,
\begin{equation}
   \hat H_b +\hat H_{sb} = 
       \frac{1}{2}\sum_{j=1}\left[ \hat p_j^2 
       + \omega_j^2\left(x_j -\frac{c_j}{\omega_j^2} f(\hat q)\right)^2 \right]
\end{equation}
Each of the bath oscillators are characterized by their respective
frequency, $\omega_j$, and coupling, $c_j$, to the system
through an arbitrary functional form, $f(\hat q)$.
The reduced density matrix at the inverse temperature, $\beta=1/k_{\rm B}T$,
is then obtained by tracing out the bath from the Boltzmann operator
of the total Hamiltonian,
\begin{equation}
   \hat \rho(\beta) = \frac{1}{Z}{\rm Tr_b}\; e^{-\beta \hat H} 
   \;.\label{eq:rho_reduced}
\end{equation}
where the partition function, $Z={\rm Tr}\left[ \exp(-\beta \hat H)\right]$.

For equilibrium quantities it is convenient to compute the reduced density
matrix in the path integral representation.
For Hamiltonians such as Eq.~\ref{eq:Ham}, 
it is well known that the trace over the harmonic bath degrees of freedom 
may be performed analytically leading to reduced equations for the 
system variables 
only.\cite{feynman63,schul86,chandler91,Weiss99,Ingold88,kleinert04}
In this case, the imaginary time path integral expression for the
elements of the reduced density matrix is given by,
\begin{equation}
   \rho(x^\prime,x;\hbar\beta) = \frac{1}{Z} \int D[q] 
   \exp\left(-\frac{1}{\hbar}\left(S_s^E[q]+\Phi[q]\right)\right)
   \;. \label{eq:path_integral}
\end{equation}
The functional integral is over all imaginary time paths, $q(\tau)$, that
satisfy the boundary conditions $q(0) = x$ and $q(\hbar \beta) = x^\prime$.
The factor $S_s^E[q]$ denotes the Euclidean action of the system,
\begin{equation}
   S_s^E[q]  = \int_0^{\hbar\beta} d\tau \frac{1}{2}M \dot q^2 + V(q)
   \;,
\end{equation}
although its explicit form will not be needed in the ensuing manipulations.
Assuming that the bath remains in thermal equilibrium, then
the trace over the bath degrees of freedom
leads to the Feynman-Vernon influence functional, $\Phi[q]$, 
in Eq.~\ref{eq:path_integral}.
This functional accounts for all of the effects of the environment
and is explicitly given by,
\begin{equation}
   \Phi[q] = -\frac{1}{2}
   \int_0^{\hbar\beta} d\tau \int_0^{\hbar\beta} d\tau^\prime
    f(q(\tau)) K(\tau - \tau^\prime)  f(q(\tau^\prime)) \;,
    \label{eq:phi}
\end{equation}
where the kernel, $K(\tau)$, is the imaginary time version of the 
force autocorrelation function of the harmonic bath,
\begin{equation}
   K(\tau) 
   = \frac{1}{\pi}\int_0^\infty d\omega \; J(\omega)  
   \frac{\cosh\left(\frac{\hbar\beta\omega}{2} - \omega\tau\right)}  
     {\sinh\left(\frac{\hbar\beta\omega}{2}\right)}
     \;, \label{eq:ff_correlation}
\end{equation}
and $J(\omega)$ is the bath spectral density,
\begin{equation}
   J(\omega) = \frac{\pi}{2}\sum_{i=1} \frac{c_j^2}{\omega_j}
   \delta(\omega - \omega_j) \;.
\end{equation}

While elegant, there are two issues with the influence functional approach 
that prevent a straightforward computation of the entire reduced density matrix.
First, due to the boundary conditions specified by Eq.~\ref{eq:path_integral}
a separate calculation must be performed for each element of the density matrix.
This can become quite time consuming for large systems.
Secondly, $K(\tau)$ generally has a long correlation time which results in a
costly time convolution integral in the influence functional, Eq.~\ref{eq:phi},
for each sampled path.
It is well known that the difficulty associated with the latter of these
issues may be alleviated, albeit at the price of introducing an additional 
functional integral.
In this approach, one applies a Gaussian integral identity commonly 
referred to as the Hubbard-Stratonovich transformation 
(uncompleting the square) to the influence 
functional in Eq.~\ref{eq:phi}.\cite{feynman63,schul86, chandler91}
In one dimension, this transformation is simply the Gaussian integral,
\begin{equation}
   \exp\left(\frac{b^2}{2a}\right)  = 
 \sqrt{\frac{a}{2\pi}}  \int dx \; \exp\left(- \frac{1}{2} a x^2 + bx\right)
   \;. \label{eq:HS}
\end{equation}
The multi-dimensional version necessary for the present case is 
presented in the final chapter of Ref.~\onlinecite{schul86}.
It underlies the auxiliary field Monte Carlo techniques\cite{sugiyama86} 
as well as similar schemes that have been recently 
proposed to calculate the dynamics in open quantum 
systems.\cite{mak91,stockburger02,shao04,makri98} 
Using this relation, the influence functional can be exactly rewritten as
\begin{equation}
   \exp\left(-\frac{1}{\hbar}\Phi[q]\right) = 
   \int D[\xi] \; W[\xi] 
   \exp\left(-\frac{1}{\hbar} 
   \int_0^{\hbar\beta} d\tau \xi(\tau)f(q(\tau))\right)
   \;, \label{eq:noise_dist}
\end{equation}
where 
\begin{equation}
    W[\xi]  = 
   \left[{\rm{det}}\left( 2\pi\hbar K(\tau-\tau^\prime)\right)\right]^{-1/2}
   \exp\left(-\frac{1}{2\hbar}\int_0^{\hbar \beta} \int_0^{\hbar\beta}
   \xi(\tau) \left[K(\tau -\tau^\prime)\right]^{-1} \xi(\tau^\prime) \right)
   \;.\label{eq:probability_distribution}
\end{equation}
Since the covariance matrix in Eq.~\ref{eq:probability_distribution}
is real and symmetric, $W[\xi]$ 
is a well-defined probability distribution.
Notice also that the costly time non-local interactions involving the system 
degrees of freedom in Eq.~\ref{eq:phi} have been exchanged for local 
interactions by introducing the additional functional integral over 
the auxiliary variable, $\xi$.
With this result the influence functional can then be combined with 
the system action remaining in Eq.~\ref{eq:path_integral}, 
so that the reduced density matrix element is given by
\begin{equation}
   \rho(x^\prime,x;\hbar\beta) = \frac{1}{Z} 
   \int D[\xi] \; W[\xi] \int D[q] 
   \exp\left(-\frac{1}{\hbar}\int_0^{\hbar\beta} 
   d\tau \frac{1}{2}M \dot q^2 + V(q) + \xi(\tau) f(q) \right)
   \;. \label{eq:noisy_pi}
\end{equation}
It is now clear that the imaginary time dynamics may be interpreted as 
one governed by a time-dependent Hamiltonian where the driving is 
determined by the $\xi(\tau)$.
The characteristics of the latter are determined by the Gaussian 
functional in Eq.~\ref{eq:probability_distribution}, 
with the covariance matrix, $K(\tau)$.
That is, $\xi(\tau)$ is colored noise which obeys the autocorrelation relation,
\begin{equation}
   \left \langle \xi(\tau) \xi(\tau^{\prime})\right \rangle = 
   \hbar K(\tau-\tau^{\prime}) \;.
\end{equation}

To clarify this result and develop a more suitable numerical scheme
for excitonic systems, it is advantageous at this point to leave the
path integral representation.
Consider the imaginary time evolution of the reduced density matrix for a 
given realization of the noise.
In this case, the Bloch equation corresponding to Eq.~\ref{eq:noisy_pi}
for the (unnormalized) density matrix is given by
\begin{equation}
   -\hbar \frac{\partial}{\partial\tau} \hat \rho = 
   \left(\hat H_s + \xi(\tau)f(\hat q) \right) \hat \rho
   \;.\label{eq:Bloch}
\end{equation}
The exact reduced density matrix is then obtained by performing the 
additional functional integral over $\xi$, which, 
in this case, corresponds to averaging over many realizations of the noise.
Each individual sample of the reduced density matrix has no physical content.
It is only after averaging over the noise that one obtains meaningful results.
The external Gaussian field is simply an efficient manner to sample the 
influence functional.
Eq.~\ref{eq:Bloch} is the main result of this section and provides
the working expression for the numerical simulations below.
Notice that this result does not rely on the particular choice of
system Hamiltonian or coupling to the bath. 
As such, it is valid for arbitrary multi-dimensional 
Hamiltonians as well as discrete systems.
The details specific to the latter are discussed in 
Appendix~\ref{sec:Discrete_PI}.
This approach is exact and generates the entire reduced density matrix from 
a single Monte Carlo calculation without any restriction
to the particular form of the spectral density $J(\omega)$.

\subsection{Numerical details}

In general the time-dependent Hamiltonians $H(\tau_1)$ and $H(\tau_2)$ 
in Eq.~\ref{eq:Bloch} do not commute. 
However, if the time step is sufficiently small  
then a symmetric Trotter expansion can be used to write
the short time propagator as
\begin{equation}
   \hat \rho(\tau + \Delta \tau) \approx 
                        \exp\left(-\frac{\Delta \tau}{2} \hat H_1(\tau) \right)
                        \exp\left(-\Delta \tau \hat H_s \right)
                        \exp\left(-\frac{\Delta \tau}{2} \hat H_1(\tau) \right)
                        \hat \rho(\tau)
                        \;, \label{eq:propagation}
\end{equation}
where $H_s$ is the bare system Hamiltonian, and $H_1$ characterizes the 
noisy system-bath interactions.
A straightforward method is used to generate a given realization of the noise.
First the covariance matrix is formed and diagonalized. 
Then the resulting independent Gaussian distributions are sampled, 
followed by a transformation back to original coordinates.

The discrete nature of the Hamiltonian for the excitonic systems analyzed 
below simplify the calculations considerably.
In this case, the bare system propagator involving $\hat H_s$ 
can be diagonalized and stored so that the central step in
Eq.~\ref{eq:propagation} can be performed exactly with a 
single matrix multiplication.
A second simplification is that the environment 
is assumed to only couple to the site populations implying that $H_1$ is 
diagonal.
As a result, the only error incurred in the propagation is due to 
the Trotter expansion which can be made arbitrarily small.
However, these two simplifications are by no means necessary in more
general cases.
For example, in continuous systems a split-operator approach or any number
of other approximate methods for the short time propagators may be employed.

As the temperature is lowered, more time slices are needed in order
to prevent the error due to the Trotter factorization in 
Eq.~\ref{eq:propagation} from becoming too large.
However, each additional time slice leads to an additional Gaussian integral 
over the noise.
This, in turn, requires a larger number of Monte Carlo samples to converge 
the functional integral in Eq.~\ref{eq:noise_dist}.
In the calculations presented below, 
convergence was obtained with between $10^4$ Monte Carlo samples at high 
temperature and weak coupling to $10^7$ samples at low temperatures and
large coupling.
The overall simplicity of this approach for calculating the reduced density
matrix makes it highly attractive compared with straightforward path integral 
implementations.

\section{Model systems}\label{sec:Models}

The single-excitation manifold of the light harvesting systems 
are modelled by the displaced oscillator Hamiltonian,
\begin{equation}
   \hat H_s = \sum_{i=1}^N E_i   \left| i \right \rangle \left \langle i \right|
       + \sum_{i\neq j}^N V_{ij} \left| i \right \rangle \left \langle j \right|
       + \sum_{i=1}^N 
          \left| i \right \rangle \left \langle i \right |
          \left[ H_b^{(i)} - c_{j}^{(i)} x_{j}^{(i)} + \eta^{(i)} \right]
       \label{eq:LH2}
       \;,
\end{equation}
where $E_i$ and $V_{ij}$ denote the site energies and couplings,
respectively. 
In recent work, Olbrich and Kleinekath{\"o}fer reported extensive molecular 
dynamics simulations of LH2 and FMO in order to characterize the 
dynamic fluctuations of the environment.\cite{olbrich10,olbrich11a}
They demonstrated that there is little spatial correlation in
the fluctuations of site energies of these systems due to the environment.
Therefore in Eq.~\ref{eq:LH2}, each site is taken to be linearly coupled 
to an independent bath of harmonic oscillators. 
The notation $x_{j}^{(i)}$ denotes the $j$-th oscillator of the bath 
that is associated with site $i$ of the system,
and $\eta^{(i)}$ its associated reorganization energy. 
In the context of Eq.~\ref{eq:Bloch}, each independent bath gives rise to 
an independent source of noise coupled to its respective site.

For FMO, we adopt the 7-site Hamiltonian determined in Ref.~\onlinecite{cho05}
through fitting to experimental spectroscopic data. 
The specific numerical values for the Hamiltonian are provided in 
Appendix~\ref{sec:FMO_ham}.
The light harvesting complex in LH2 consists of $N=18$ 
chromophores arranged in a ring of $9$ dimers.
As a result of the dimerized structure, the site energies take on the
alternating values of $12, 458$ cm$^{-1}$ and $12, 654$ cm$^{-1}$.
The nearest neighbors couplings within a dimer are
by $363$ cm$^{-1}$ and the nearest neighbor inter-dimer couplings 
are $320$ cm$^{-1}$.\cite{strumpfer09}
The remaining non-nearest neighbor couplings are assumed
to be given by dipole-dipole interactions determined from
\begin{equation}
   V_{ij} = C \left(\frac{\mathbf d_i \cdot \mathbf d_j}
                   {\left| \mathbf r_{ij}\right|^3}
   -3 \frac{\left(\mathbf d_i \cdot \mathbf r_{ij}\right) 
            \left(\mathbf d_j \cdot \mathbf r_{ij}\right)  }
           {\left| \mathbf r_{ij}\right |^5} \right)
           \;,
\end{equation}
where the constant $C=348000$ \AA$^3$ cm$^{-1}$.
The geometry of LH2 needed for constructing the dipole-dipole
interactions is taken from the crystal structure and follows the 
prescription of Ref.~\onlinecite{damjanovic02}.

In the simulations of LH2 by Olbrich and Kleinekath{\"o}fer, they
were also able to extract the time correlations functions of the 
fluctuations in the site energies that arise from the system bath 
coupling.\cite{olbrich10}
After fitting the correlation function to a simple analytic form, they
provide the following estimate for the spectral density of the bath,
\begin{equation}
   J(\omega) = \frac{2}{\hbar} \tanh(\hbar \beta \omega/2) \left(
    \sum_{i=1}^2 \frac{\eta_i \omega_{c_i}}{\omega^2 + \omega_{c_i}^2}
  + \sum_{i=1}^{10} \frac{\bar\eta_i \bar \omega_{c_i}}
                    {(\omega - \bar\omega_i)^2 + \bar\omega_{c_i}^2}
  \right)
  \;.
   \label{eq:sd_olbrich}
\end{equation}
The temperature dependent prefactor reflects the nonlinear nature
of the bath in this model.\cite{Weiss99}
The relevant parameters for both the B850 and B800 rings of LH2 are listed in
Ref.~\onlinecite{olbrich10}.
We assume for both LH2 and FMO that the spectral densities of the independent
baths are identical.

Rather than taking such a detailed approach in describing the bath, 
however, most studies of light harvesting systems often assume the bath
spectral density has a simple Ohmic form with a Lorentzian or exponential
cutoff.\cite{ishizaki09a, strumpfer09} 
In order assess the influence of the reorganization energy on the coherence 
length we will also use the Ohmic spectral density,
\begin{equation}
   J(\omega) = \frac{\pi \eta \omega}{\hbar \omega_c} e^{-\omega/\omega_{\rm c}}
   \label{eq:sd_ohmic}
   \;.
\end{equation}
For comparison, the parameters for FMO reported in Ref.~\onlinecite{ishizaki09a}
are $\eta=35$ cm$^{-1}$ with a cutoff frequency of $0.02$ fs$^{-1}$.
For LH2, the values of $\eta=200$ cm$^{-1}$ and  $\omega_c = 0.01$ fs$^{-1}$ 
are employed in Ref.~\onlinecite{strumpfer09}.
However, somewhat  larger values for both the reorganization energy and
disorder have also been used to fit experimental 
data,\cite{kuhn97, novoderezhkin98}, as well as in studies of artificial
circular excitonic systems.\cite{donehue11}
For simplicity, we assume that each bath is characterized by the 
same spectral density.

In addition to the homogeneous broadening in the excitonic systems
due to the bath, a substantial amount of inhomogeneous broadening 
is also present. 
In the simulations below, the latter is accounted for by introducing 
a source of static disorder on the site energies.
For simplicity we assume that the energies of each site are broadened
by an independent Gaussian distribution
\begin{equation}
   P_i(\epsilon) = \frac{1}{\sqrt{2\pi\sigma_i^2}} 
   e^{-(\epsilon-E_i)^2/2\sigma_i^2} 
   \;. \label{eq:disorder}
\end{equation}
The distribution is assumed to be the same at each site so that
all of the variances are equal, $\sigma_i^2 = \sigma^2$. 
The static disorder has been estimated to be $80$ cm$^{-1}$ in FMO 
and $200$ cm$^{-1}$ in LH2.\cite{milder10,strumpfer09}

Many of the qualitative features of the effects of the environment on
the coherence length in FMO and LH2 may be explained in terms of the 
simple two level system with static disorder.
Therefore results are first presented for the two level system,
\begin{equation}
   \hat H_s = J \hat \sigma_x + \Delta \hat \sigma_z\;, 
\end{equation}
where $\sigma_x$ and $\sigma_z$ are the respective Pauli matrices.
The detuning of the energy levels is governed by $\Delta$ and the coupling
between the two sites is denoted by $J$.

\subsection{Coherence Length Measures}

The coherence length referred to in light harvesting systems is a measure
of the extent of the off-diagonal elements of the reduced density matrix 
in the site basis.
This quantity has no precise definition so below we shall consider
two different commonly used proxies for the coherence length.
The first of which is defined by\cite{meier97,dahlbom01}
\begin{equation}
   L_{\rho}=\frac{\left(\sum_{ij}^N \left| \rho_{ij}\right| \right)^2}
                {N \sum_{ij}^N \left| \rho_{ij}\right|^2}
                \label{eq:IPR}\;.
\end{equation}
This function is essentially a measure of the variance of the density matrix
and has a direct relationship to the superradiance enhancement 
factor.\cite{meier97}
For the ensuing discussion, it is useful to analyze the limiting behavior
of $L_{\rho}$.
In the high temperature limit, the density matrix 
describes an incoherent superposition of states that 
is diagonal with equal populations on all the sites.
There are no coherences in this case so the density matrix 
reduces to $\rho_{ij} = \frac{1}{N}\delta_{ij}$ and $L_{\rho}=1$.
In the opposite limit of complete coherence, all of the elements 
$\rho_{ij}=1/N$ and $L_{\rho}=N$.
Finally, in the case of a pure state, the density matrix is again diagonal
except that all of the population is localized at a single site.
In this case, $L_{\rho}=1/N$.

As another definition for the coherence length, one may use the alternative 
construct\cite{kuhn97,dijkstra08}
\begin{equation}
   L_{c} = \sum_{m=1}^N \sum_{n=m}^N \left| p_{m,m+n} \right|
   \label{eq:L}\;.
\end{equation}
In this case, if the density matrix is completely coherent 
so that $\rho_{ij} = 1/N$, then $L_c$ reaches its maximum value of $(N+1)/2$.
For both a diagonal density matrix and a pure state, $L_c={\rm Tr} \rho=1$. 
That is, this definition does not distinguish between a density matrix 
that describes a pure state and one that characterizes an
incoherent superposition of states.
The behavior of both of these definitions, as well as several others,
has been discussed at length in Ref.~\onlinecite{dahlbom01}
in the specific context of LH2 with static disorder.
Below, we will not be particularly interested in the absolute values
of the either definition, only the ability of the respective coherence
length to accurately reflect the changes in the density matrix induced by 
the environment.

\section{Numerical results} \label{sec:Results}

\subsection{Two level systems}
\label{sec:TLS}

It will become apparent below that many of the qualitative features 
seen in the coherence length of the two more complicated systems 
can be captured by the respective symmetric and biased systems.
Before discussing numerical results, it is helpful to analyze the 
limiting cases of the parameters in this simple model denoted by 
the variance of the static disorder, $\sigma^2$, 
inverse temperature $\beta$, coupling $J$, and bias, $\Delta$. 
The reduced density matrix for the two level system without
disorder is easily given by
\begin{equation}
   \hat \rho(\beta) =
       \frac{1}{2\lambda}\left( \lambda \hat I - 
       \tanh(\beta\lambda) \hat H \right)
       \;,
\end{equation}
where $\hat I$ is the identity matrix and the eigenvalue 
$\lambda= \sqrt{J^2 + \Delta^2}$.
For this case the coherence length 
$L_c = 1 + \frac{J}{2\lambda}\tanh(\beta\lambda)$,
whereas
\begin{equation}
   L_{\rho} = \frac{\left( 1 + \frac{J}{\lambda}\tanh(\beta\lambda)\right)^2}
                  {1 + \tanh^2(\beta\lambda)}
                  \;.\label{eq:lrho_tls}
\end{equation}
In the high temperature limit, the reduced density
matrix represents a classical mixture regardless of the values 
of the other parameters and both of the coherence lengths are $1$.
In the limit of finite temperature but no static disorder,
then the off-diagonal matrix elements are given by $\tanh(\beta \lambda)/2$,
and at sufficiently low temperature, all of the elements of the 
reduced density matrix reach a (maximal) value of $1/2$ indicating 
a completely coherent state.
As the width of the static disorder is increased, the fluctuations will 
eventually destroy the coherences even at low temperature.
For the symmetric two level system, 
each realization of the static disorder leads to a detuning of 
the energy levels and thus to decrease in the coherence length.
When the width of the disorder distribution is sufficiently large, 
the averaged density matrix is again diagonal but the mechanism 
is different than in the high temperature limit.
In this case each realization of the disorder leads to a 
density matrix that is localized on one of the two sites.
It is only after averaging over the distribution of static disorder that
one recovers a density matrix with equal populations.

\begin{figure}
   \includegraphics*[width=0.85\textwidth]{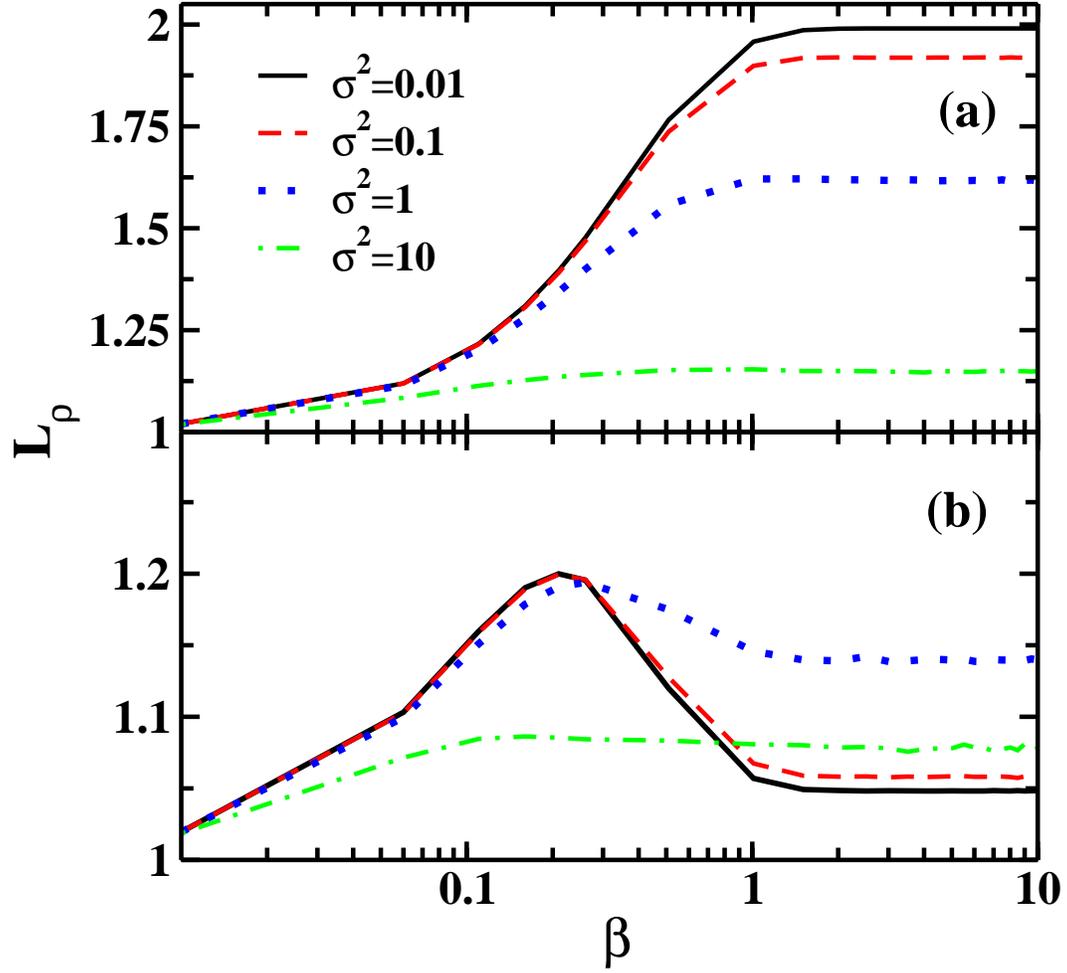}
   \caption{(color online)
   The effect of static disorder on $L_{\rho}$
   calculated for the symmetric (top) and biased (bottom) 
   two level system.
   In both cases $J=1$ and the detuning is $\Delta = 2J$ biased case.
   The variance of the static disorder is 
   $\sigma^2=10$ for the (green) dot-dashed line,
   $1$ (blue) dotted line, $0.1$ (red) dashed line and $0.01$ (black) solid
   line.
   Note the plateau at low temperatures in the symmetric case and the
   peak in (b).
   }\label{fig:TLS_IPR}
\end{figure}

To make these claims more concrete, the coherence length $L_{\rho}$ 
is shown for the symmetric and biased two level system in 
Fig.~\ref{fig:TLS_IPR}(a) as a function of the inverse temperature
for various values of the static disorder.
As described above, for the symmetric case with very weak static disorder, 
$L_{\rho}$ increases as the temperature is lowered eventually reaching
the maximal value of $2$.
As the width of the static disorder distribution increases, the 
coherence length is steadily reduced.
That is, the unbiased case is maximally coherent without disorder.
Static disorder can only lead to a decrease of the coherence length
in symmetric systems.

While the biased two level system behaves similarly at high
temperatures, it displays qualitatively different behavior in the 
low temperature regime as shown in Fig.~\ref{fig:TLS_IPR}(b).
There are two interesting features in this case. 
Most notably, a maximum in $L_{\rho}$ appears as a function of the 
temperature even in the absence of disorder. 
Secondly, it can be seen that a finite amount of disorder can increase
the coherence length at low temperature.
The former feature may be explained by noting that 
a maximum in $L_{\rho}$ will be observed when the derivative of 
Eq.~\ref{eq:lrho_tls} with respect to $\beta$ is zero.
This leads to the the explicit relation
\begin{equation}
   \beta = \frac{1}{2\lambda}\ln\left(\frac{\lambda + J}{\lambda-J}\right)
   \;,\label{eq:max}
\end{equation}
which may be expressed alternatively in terms of the 
partition function as $\frac{\partial\ln Z}{\partial \beta} = J$.
For the symmetric two level system ($\lambda = J$), 
Eq.~\ref{eq:max} demonstrates that a maximum exists only a zero temperature.
As the bias increases the maximum shifts to higher temperatures.
This feature, however, is present only for this particular definition of 
coherence length.
No such maximum can exist for $L_c$; this measure increases 
monotonically with the inverse temperature.
As noted above $L_{\rho}$ encodes information about the populations as well 
as the coherences while $L_c$ does not.
The peak in $L_{\rho}$ along with values less that $1$ are an indication 
that the population is becoming localized on one of the sites.

\begin{figure}
   \includegraphics*[width=0.85\textwidth]{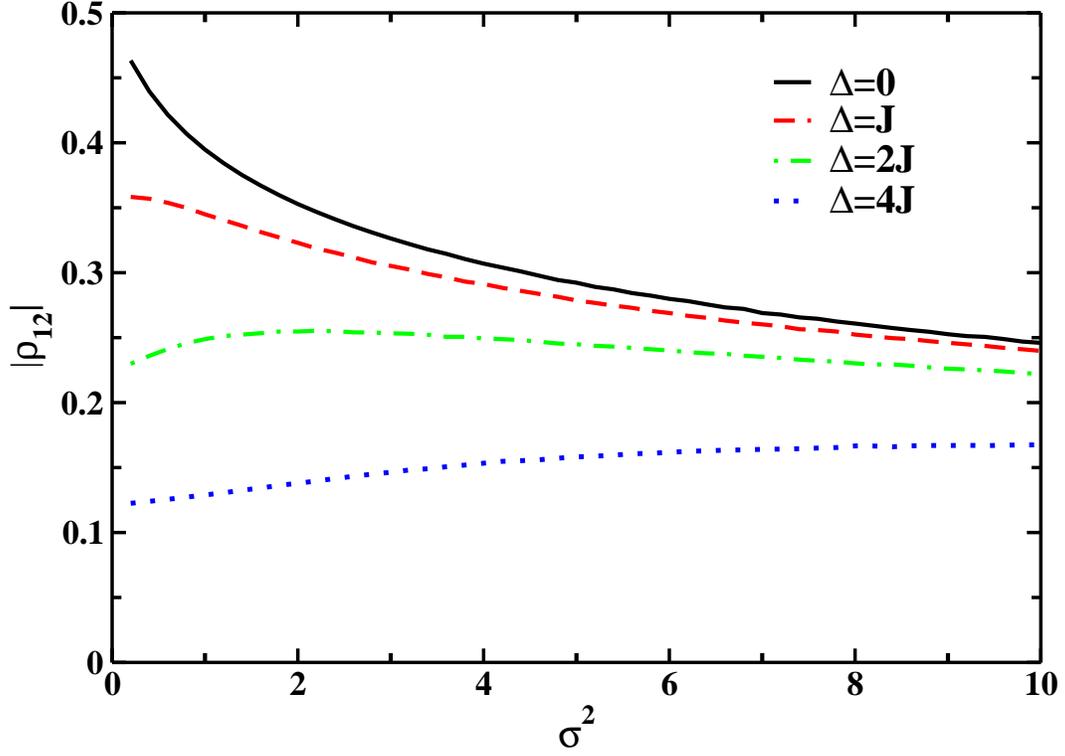}
   \caption{(color online)
   The off diagonal element of the density matrix calculated as a
   function of the variance of the static disorder distribution.
   In both cases $J=1$ and the temperature is fixed at $\beta=10$.
   The (green) dot-dashed line, (blue) dotted line, (red) dashed line,  
   and (black) solid line correspond to detunings
   of the energy levels of $\Delta=4J$, $2J$, $J$, $0$, respectively.
   }\label{fig:TLS_p12}
\end{figure}

The second interesting feature of Fig.~\ref{fig:TLS_IPR}(b) is that 
at low temperatures, increasing the width of static disorder may
lead to an increase of $L_{\rho}$.
This behavior is captured by either definition of the coherence length.
For the two-level system, the coherence length is determined simply by the 
single off-diagonal element of the reduced density matrix.
In Fig.~\ref{fig:TLS_p12}, $\rho_{12}$
is shown as a function of the width of the static disorder distribution 
for various cases of the bias with a fixed inverse temperature of $\beta=10$.
This temperature is sufficiently low such that it is only necessary to 
consider the off-diagonal element of the reduced density matrix at zero
temperature.
In this case, the disorder-averaged density matrix element is given by 
\begin{equation}
   \left| \rho_{12} \right| = \frac{J}{2}\left 
   \langle\left[J^2 + (\Delta + \epsilon)^2\right]^{-1/2} \right\rangle
   \;,\label{eq:TLS_disorder_avg}
\end{equation}
where the average is taken over the distribution of Gaussian disorder,
$P(\epsilon)$ in Eq.~\ref{eq:disorder}
Only for the case of the symmetric two level system
can $\rho_{12}$ be calculated exactly for which one obtains,
\begin{equation}
   \left| \rho_{12} \right| = \frac{1}{\sqrt{2 \pi \sigma^2} }
                              e^{\frac{J^2}{4 \sigma^2}}\frac{J}{2}
                              K_0\left( \frac{J^2}{4 \sigma^2}\right)
                              \;, \label{eq:rho00}
\end{equation}
where $K_0$ is the zero-order modified Bessel function of the second kind.
This expression decreases monotonically with increasing width of the 
static disorder distribution as also seen from Fig.~\ref{fig:TLS_IPR}(a).
That is, the introduction of disorder can only decrease the coherence
length in the symmetric two level system.

Likewise, the two level system is maximally coherent also in the absence 
of bias.
The presence of bias reduces $\rho_{12}$ by a factor of 
$J/\sqrt{J^2 + \Delta^2}$ at low temperature.
Provided that the ratio $J^2/\sigma^2$ is large, 
the initial increase in the coherence length of the biased two level
system with increasing disorder seen in Figs.~\ref{fig:TLS_p12} 
and~\ref{fig:TLS_IPR}(b) may be computed from Eq.~\ref{eq:TLS_disorder_avg} as
\begin{equation}
   \left| \rho_{12}\right| \approx \frac{J}{2\lambda}
                        \left(1 + \sigma^2\frac{2\Delta^2 - J^2}{2\lambda^4}
                        + O(\sigma^4)\right)\;.
\end{equation}
As can be seen from this expression, for $2\Delta^2 > J^2$, 
there will be an initial increase in the magnitude of 
the off-diagonal element of the reduced density matrix, and hence
also the coherence length.

The physical origin of the increase in the coherence length may
be explained by realizing that the disorder gives rise to a 
distribution of site energies.
The coherence length will be largest when the overlap of the two 
site energy distributions is maximized.
It is easily seen for the symmetric two level system that this criterion 
implies that the coherence is largest only when the distribution of static 
disorder has zero width as seen in 
Figs.~\ref{fig:TLS_IPR}(a) and~\ref{fig:TLS_p12}.
However for the biased case, there will be an optimal width of the static
disorder distribution that will increase the overlap of the two sites.
This leads to a maximum in the coherence length at a
non-zero value of the width of the static disorder distribution in biased
systems as seen in Fig.~\ref{fig:TLS_IPR}(b).
It will be seen below that many of these simple qualitative considerations
regarding the coherence length in two level systems will also hold true
in the more complicated settings of LH2 and FMO.

\subsection{LH2}
\subsubsection{Reduced density matrix}

\begin{figure}
   \includegraphics*[width=0.45\textwidth]{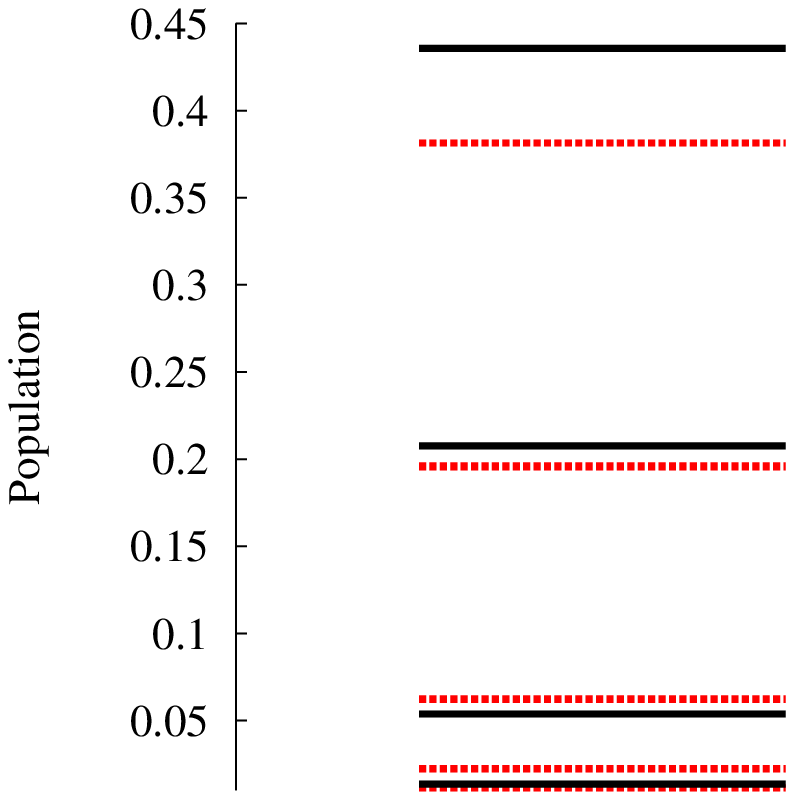}
    \caption{(color online)
    The Boltzmann populations (black) lines and the exact equilibrium 
    populations (red) lines for the seven most populated states of LH2
    at $300$ K.
    The lower three states are doubly degenerate.
    The thermal bath is modeled by the Ohmic spectral density described 
    in Eq.~\ref{eq:sd_ohmic} with a reorganization energy of 
    $200$ cm$^{-1}$ and cutoff frequency of $100$ fs$^{-1}$.
   }
   \label{fig:populations}
\end{figure}

In Ref.~\onlinecite{strumpfer09}, Str{\"u}mpfer and Schulten presented the 
numerically exact time evolution of the exciton populations in LH2 obtained 
from the hierarchical equation of motion approach.
They noted that the long time, steady state behavior of the exciton 
populations did not coincide with the those calculated from the
Boltzmann populations of the bare system Hamiltonian.
The difference is not entirely negligible; it lowers the population of the 
most lowest energy state by roughly $15\%$.
Here we show that the steady state that is reached in the long time limit of 
their calculations is simply the true equilibrium state of the 
full system-bath Hamiltonian.
In Fig.~\ref{fig:populations}, the Boltzmann populations of the seven
most populated exciton states (the lower three are doubly degenerate)
are presented as well as the corresponding values obtained from the 
exact path integral calculations.  
The latter quantities are given by $P_i(\beta) = \left \langle \psi_i \right 
| \rho(\beta) \left | \psi_i \right\rangle$, where $\psi_i$ labels 
the $i$-th exciton basis function.
The results in Fig.~\ref{fig:populations} 
are in excellent agreement with corresponding values presented in 
Ref.~\onlinecite{strumpfer09}.
The difference between the populations calculated from the Boltzmann 
distribution and the reduced density matrix seen in Fig.~\ref{fig:populations}
is an indication of the entanglement of the system and bath in 
the single excitation manifold.
That is, the true equilibrium state for LH2 
cannot be written as a product of independent system and bath states 
as is often assumed, even at room temperature.
For example, in subsequent calculations of the energy transfer rate between two
LH2 rings based up Forster theory, Str{\"u}mpfer and Schulten 
noted that using the correct equilibrium distribution instead of the 
populations given by the Boltzmann distribution leads to a corresponding 
decrease in the transfer rate of approximately $10\%$.
As will be demonstrated below, this correction becomes more important 
when either the temperature is lowered or the system-bath coupling increases.

\begin{figure}
   \includegraphics*[width=\textwidth]{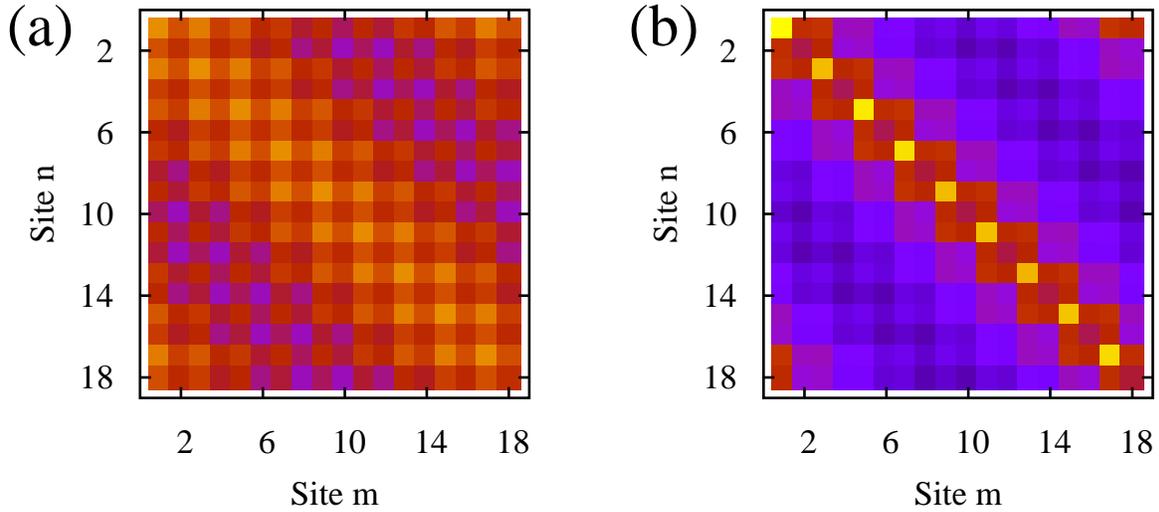}
   \caption{(color online)
   The absolute value of the density matrix elements 
   $\left|\rho_{nm}\right|$ for LH2 at a temperature of $100$ K.
   The left is the Boltzmann distribution and the right
   is the exact reduced density matrix calculated at 
   a reorganization energy of $600$ cm$^{-1}$.
   }
   \label{fig:density_matrices}
\end{figure}

In order to analyze the role of the environment in more detail, 
a comparison of the exact reduced density matrix and the approximate
Boltzmann distribution at $100$ K is shown Fig.~\ref{fig:density_matrices}.
In the absence of the thermal bath, the density matrix 
shown in Fig.~\ref{fig:density_matrices}(a) is almost completely delocalized.
However, when the bath is included in Fig.~\ref{fig:density_matrices}(b), then
the coherence length of the reduced density matrix is drastically reduced.
Due to the circular arrangement of the chromophores in LH2, 
the Boltzmann density matrix must reflect the underlying symmetry of 
the Hamiltonian.
The exact reduced density matrix must also preserve this 
symmetry since each of the independent thermal baths are identical. 
One important consequence of this result, for example, is that the exact 
reduced density matrix is diagonal in the exciton basis for any values
of the temperature or bath parameters.
An additional consequence of this symmetry is that all of the independent 
information contained in the density matrix for LH2 is captured by the 
elements of a single row or column.
Therefore, these elements will be used to provide a more quantitative 
comparison for how the temperature and system bath coupling strength 
effect the localization length in LH2.

\begin{figure}
   \includegraphics*[width=0.85\textwidth]{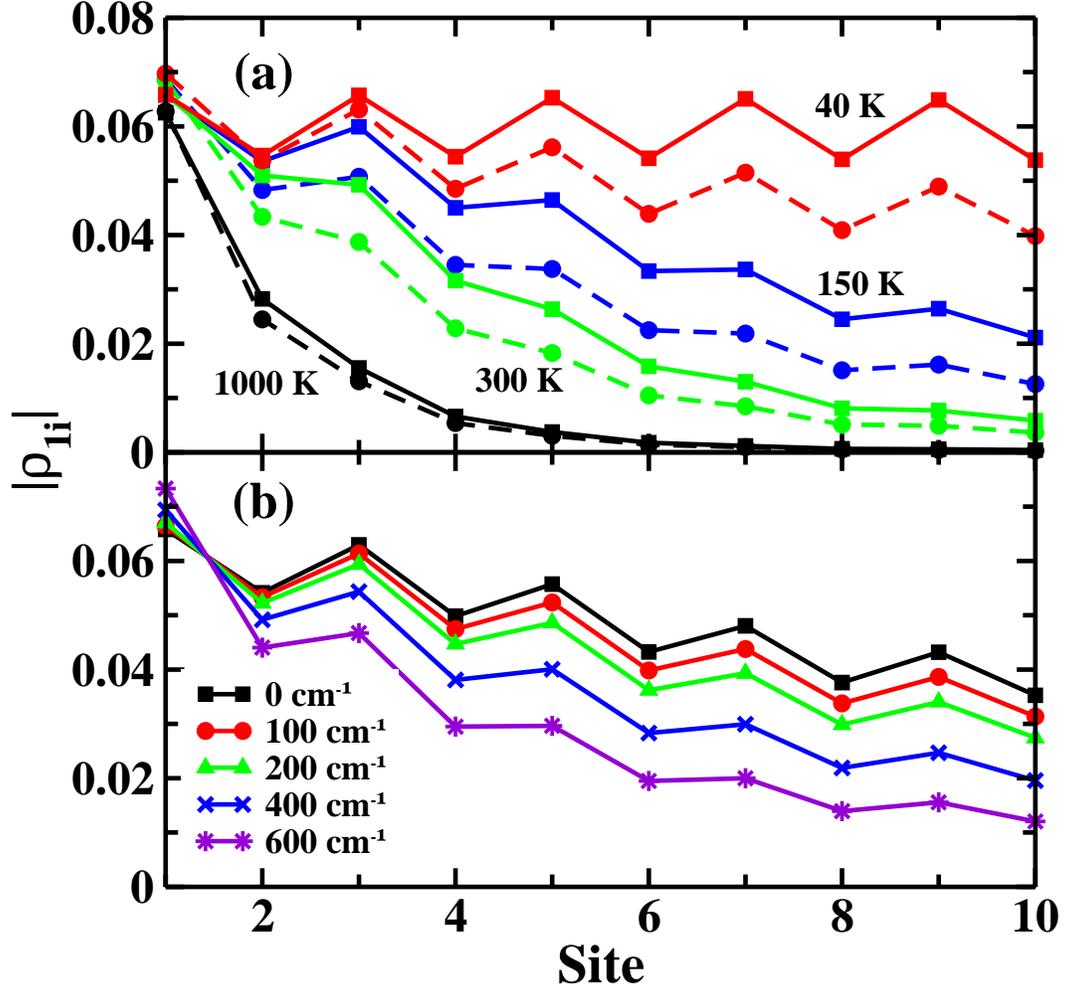} 
   \caption{(color online)
   The density matrix elements $\rho_{1i}$ for LH2 versus the site number.
   In (a) the black, green, blue and and red lines correspond to 
   temperatures of $1000$ K, $300$ K, $150$ K, and $40$ K with
   a fixed reorganization energy of $350$ cm$^{-1}$.
   The solid lines with squares are the values calculated from the
   Boltzmann distribution and the dashed lines with circles are the
   corresponding exact results from the path integral calculations.
   In (b) the (black) squares, (red) circles, (green) triangles, (blue) crosses
   and (purple) stars correspond to the Boltzmann distribution, 
   and reorganization energies of $100$, $200$, $400$ and $600$ cm$^{-1}$,
   respectively, with a fixed temperature of $100$ K.
   }
   \label{fig:rho1i}
\end{figure}

The first $10$ elements of the first row in the reduced density matrix 
are shown in Fig.~\ref{fig:rho1i}(a) for four different temperatures 
with a constant reorganization energy of $350$ cm$^{-1}$ using the Ohmic
spectral density.
Perhaps surprisingly, at $1000$ K there is still some coherence present in LH2 
as well as noticeable corrections to the Boltzmann distribution due to the bath.
As the temperature is lowered, the density matrix becomes more delocalized
as expected.
However the corrections to the Boltzmann distribution due to the bath
become more significant as well.
Notice also that the difference between the Boltzmann results 
and the exact results is not systematic with increasing distance from
the diagonal.
For example, the largest difference between the Boltzmann and exact results
shifts to larger site numbers as the temperature decreases.
This fact prevents one from representing the exact density matrix
simply as a Boltzmann distribution of the bare system with an effective 
temperature. 
The influence of the reorganization energy on the density matrix is shown 
in Fig.~\ref{fig:rho1i}(b) at a fixed temperature of $100$ K.
As the system bath coupling increases, the off-diagonal elements of 
the density matrix display a corresponding decrease.
These two results in Fig~\ref{fig:rho1i}(a) and (b) demonstrate that 
the system and bath are substantially entangled in the single exciton
manifold for almost all physically 
relevant values of the environmental parameters.
Treating the initial density matrix as a product state, as in generally
done in most calculations of the dynamics in light harvesting systems,
introduces an additional source of error that is not negligible, 
particularly at low temperatures.

\subsubsection{Coherence lengths in LH2}

\begin{figure}
   \includegraphics*[width=0.85\textwidth]{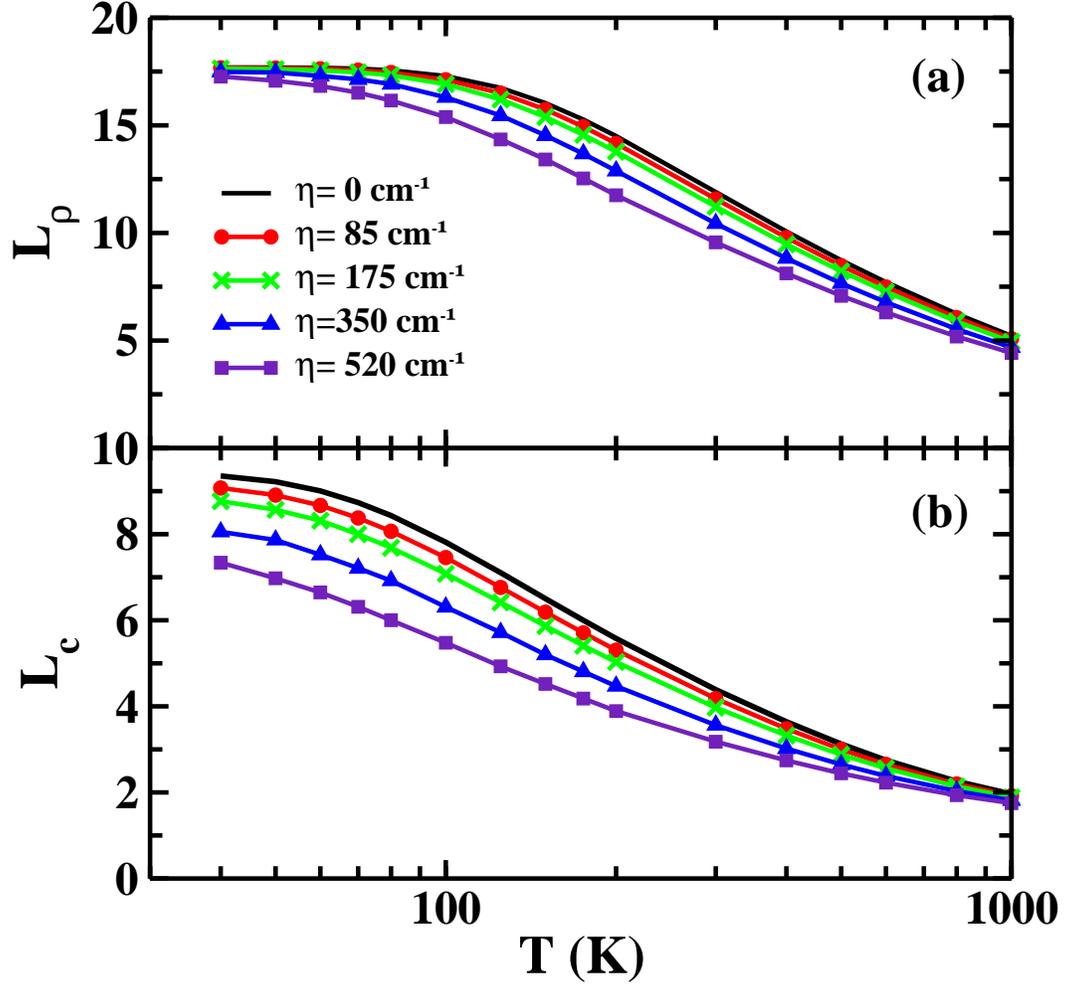}
   \caption{(color online)
    The coherence lengths, $L_{\rho}$ (a), and $L_c$ (b), 
    calculated for LH2 as a
    function of temperature for increasing reorganization energy.
    The solid (black) line is obtained from the Boltzmann populations 
    without noise.
    The reorganization energies of $85$, $175$, $350$, and $520$ cm$^{-1}$
    are indicated by the (red) line with dots, (green) line with crosses, 
    (blue) line with triangles, and (purple) line with squares, respectively.
    Note the increasing decay from the plateau at low temperatures.
   }
   \label{fig:LH2_noise}
\end{figure}

While direct inspection of the density matrix provides the most unambiguous
interpretation of the influence of the environment on the coherence length, 
it can become cumbersome for systems that lack the symmetry of LH2.
As a result a variety of measures have been proposed to quantify the 
coherence length.
They all provide some representation of the extent of the off-diagonal
elements by assigning a single number for a given density matrix.
It can already be seen from the above discussions that this distillation
of information in the density matrix cannot be completely satisfactory.
In this section we will compare two commonly used definitions for the
coherence length.
The influence of noise on the coherence length in LH2 is shown in 
Fig.~\ref{fig:LH2_noise} as a function of the temperature.
The Ohmic spectral density is used here to assess how 
the reorganization energy effects the coherence length.
Results for the spectral density of Ref.~\ref{eq:sd_olbrich} are
included below in Fig.~\ref{fig:LH2_noise_disorder}.
In the incoherent high temperature limit, the coherence length is 1
as expected since the density matrix is diagonal in this case.
However, this limit is not reached until unphysically high values of
the temperature.
As was noted from the direct examination of the density matrices above 
in Fig.~\ref{fig:rho1i}, as the temperature is lowered in the absence of noise,
the coherence length gradually increases eventually 
reaching the respective maximal value in either measure.
Likewise, increasing the strength of the system bath coupling generally 
leads to a decrease in the coherence length.
There are, however, some significant differences between the two coherence
length measures, particularly at low temperatures.
In Fig.~\ref{fig:LH2_noise}, $L_{\rho}$ shows very little dependence on the
reorganization below $100$ K regardless of the strength of the noise, 
whereas $L_c$ steadily decreases with increasing reorganization energy.
Compared with Fig.~\ref{fig:rho1i}, $L_c$ seems to provide a more consistent 
reflection of the density matrix in this case.

\begin{figure}
   \includegraphics*[width=0.85\textwidth]{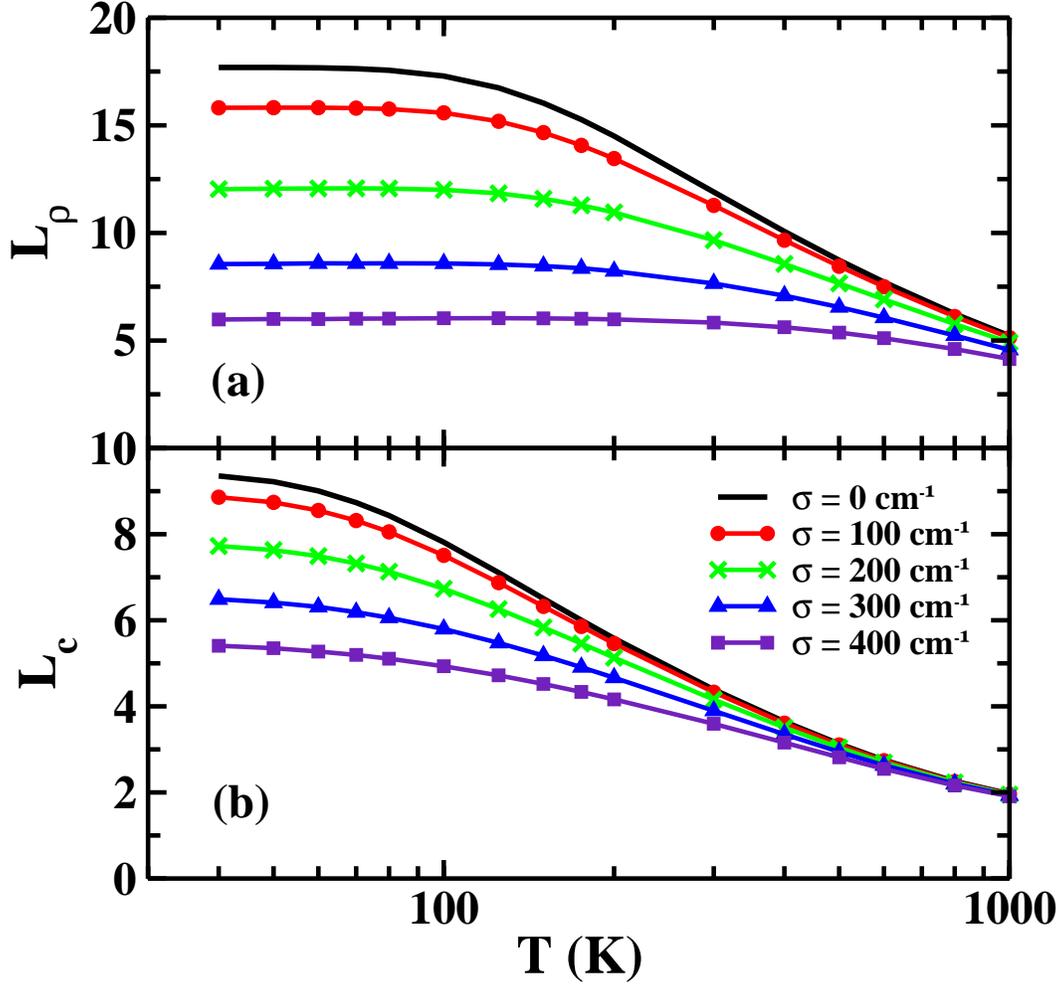}
   \caption{(color online)
    The coherence lengths, $L_{\rho}$ (a), and, $L_c$ (b), 
    calculated for LH2 as a function of temperature for 
    increasing static disorder.
    The solid (black) line is obtained from the Boltzmann populations 
    without any static disorder.
    The widths of the static disorder distribution of
    $100$, $200$, $300$, and $400$ cm$^{-1}$
    are indicated by the 
    (red) line with dots, (green) line with crosses, 
    (blue) line with triangles, and (purple) line with squares, respectively.
    Note the increasing persistence of the plateau with increasing disorder.
   }
   \label{fig:LH2_coherences}
\end{figure}

Aside from the role of homogeneous broadening, there is an additional 
source of decoherence in excitonic systems due to static disorder.
The coherence lengths calculated for LH2 with static disorder are shown in 
Fig.~\ref{fig:LH2_coherences}.
The width of the static disorder distribution was estimated to be $200$ 
cm$^{-1}$ in Ref.~\onlinecite{strumpfer09}, 
although estimates for this quantity vary widely.
As static disorder is introduced into the system, the coherence length
steadily decreases as expected from the above discussion for two level
systems.
Additionally the two different definitions for the coherence 
length display qualitatively similar behavior as has been observed 
before.\cite{dahlbom01}
However, in contrast to the case of noise, at low temperatures $L_{\rho}$ 
now displays a stronger dependence on the disorder than $L_c$.
Additionally, in both cases the coherence length reaches a plateau 
at low temperature.
Such a feature is not formed in the case of the noise.

The affect of both noise and disorder 
on the coherence length in LH2 is shown in Fig.~\ref{fig:LH2_noise_disorder}.
Here the spectral density of the bath is taken as the
form suggested by Olbrich et al in Eq.~\ref{eq:sd_olbrich}.\cite{olbrich10}
Qualitatively, the differences in the coherence lengths computed with 
this complicated spectral density and the simpler form used previously 
in Fig.~\ref{fig:LH2_noise} is rather small.
The use of Eq.~\ref{eq:sd_olbrich} leads to results 
that are quite similar to the Ohmic form used previously with a reorganization 
energy of between $200$ and $300$ cm$^{-1}$.
The spectral density of Eq.~\ref{eq:sd_olbrich} possess several strong peaks 
in the low frequency region which have been claimed to be essential in order 
to correctly describe the energy relaxation 
process in LH2.\cite{adolphs06, van_grondelle06, olbrich10}
Their influence on the coherence length, however, appears to be rather small.
The real-time dynamics are more likely to exhibit a greater sensitivity 
to these components of the respective spectral densities.
Results showing the effects of static disorder with a width 
of $200$ cm$^{-1}$ as was suggested in Ref.~\onlinecite{strumpfer09}
are also shown in Fig.~\ref{fig:LH2_coherences}.
It is important to note that the combined effect of static disorder 
and thermal noise on the coherence length is not simply cumulative.
This is particularly noticeable at low temperature in the case of $L_{\rho}$, 
but also for $L_c$.
That is, both of these effects need to be properly accounted for in this regime
in order to accurately describe the system.

\begin{figure}
   \includegraphics*[width=0.85\textwidth]{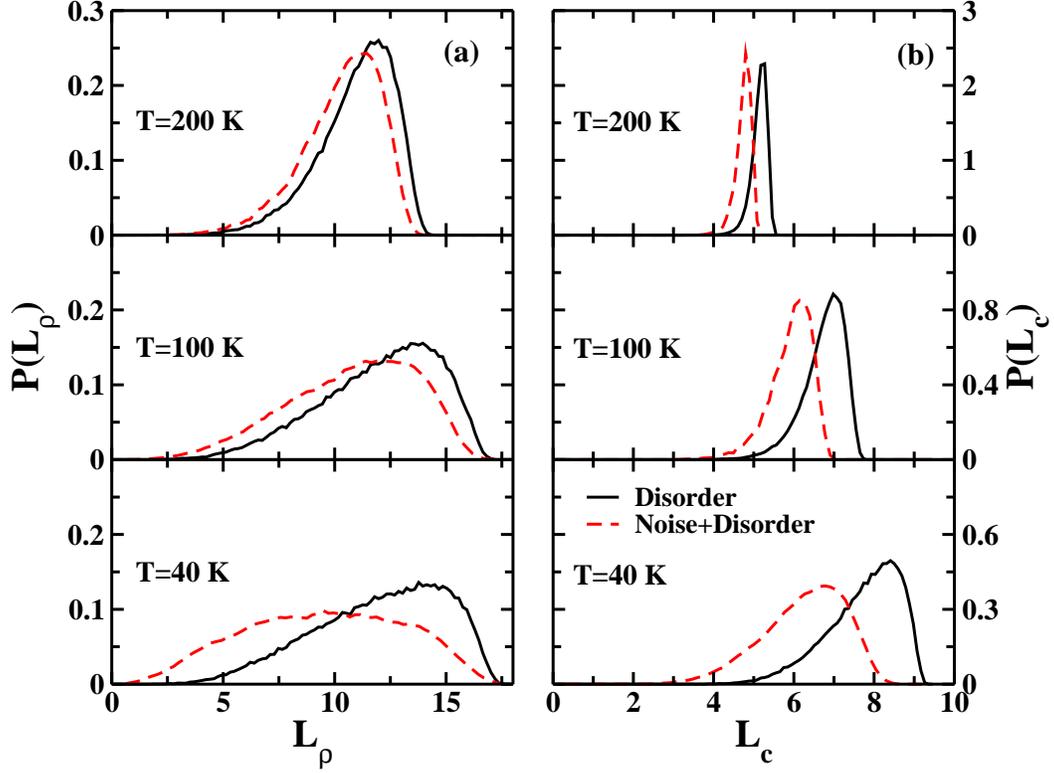}
   \caption{(color online)
    The distribution of the the coherence lengths, $L_{\rho}$ (a), 
    and $L_c$ (b) for LH2. 
    The top, middle and bottom panels correspond to temperatures of 
    $200$, $100$ and $40$ K.
    The solid (black) lines represent the distributions obtained with 
    static disorder only with a width of $200$ cm$^{-1}$.
    The dashed (red) lines display the results of disorder and noise
    using the spectral density of Eq.~\ref{eq:sd_olbrich}. 
    Note the broadening induced by disorder and the shift that results
    from the noise.
   }
   \label{fig:LH2_hist}
\end{figure}

To understand more clearly the difference between noise and disorder,
the distribution of coherence lengths calculated with disorder alone,
and with noise and disorder is shown in Fig.~\ref{fig:LH2_hist} at three
different temperatures.
The role of temperature, in general,  provides the largest contribution
to the coherence length distributions.
At high temperatures, particularly for $L_c$, the distributions are 
quite sharp and neither disorder nor noise provide a significant contribution
to the coherence length in this case.
This is consistent with the small spread in coherence lengths seen above in 
Figs.~\ref{fig:LH2_coherences} and~\ref{fig:LH2_noise} at high
temperature for any values of the reorganization energy or disorder.
The influence of static disorder is seen to both broaden the coherence 
length distributions as well as to introduce a skew towards lower values.
These features become more pronounces at low temperatures.
Including noise provides an additional constant shift of the whole 
coherence length distribution.
The notable exception to this rule is the distribution of the $L_{\rho}$ 
at $40$ K where the presence of noise causes both a large shift and 
substantial change in the shape of the coherence length distribution.
It is this effect which leads to the maximum as a function of temperature
in Fig.~\ref{fig:LH2_noise_disorder}.
One also notes that there is little difference between the static disorder
distributions of $L_{\rho}$ at $100$ and $200$ K.
This accounts for the plateau in the coherence length distributions 
seen at low temperature in Fig.~\ref{fig:LH2_coherences}.

\subsubsection{Quenched and Annealed Disorder}

Often in studies of light-harvesting systems, static
disorder is said to arise from the very low frequency motions of the bath
caused by large scale motions of the protein environment.
However, this statement in inconsistent with the manner
in which the calculations of static disorder are actually performed. 
Static disorder is due to the different local environments surrounding
each of the chromophores.
It leads different realizations of independent system Hamiltonians
as occur, for example, in impurity models. 
The disorder that results from a very slow (adiabatic) bath, 
on the other hand, arises from internal degrees of freedom.
The latter is a manifestation of annealed disorder, while the former
is known as quenched disorder.
It is generally accepted that the disorder present in light harvesting systems 
corresponds to the latter.
Each sample of the static disorder corresponds to a physically 
realizable Hamiltonian as demonstrated by single molecule experiments. 
While the subtle difference between the two may seem only semantic, 
the consequences of this distinction can by quite significant.
For example, annealed disorder implies that the system is ergodic 
and will (eventually) explore all possible bath configurations whereas quenched
disorder corresponds to a system that is not ergodic.

An adiabatic bath can be easily considered within the formalism developed
in Sec.~\ref{sec:Methods}. 
This regime is obtained in the limit that the cutoff frequency of spectral
density goes to zero.
In this case, the dynamics of the bath degrees of freedom are much slower 
than that of the system. 
The kernel $K(\tau)$ in Eq.~\ref{eq:ff_correlation} then
becomes independent of $\tau$ and may be replaced by a constant.
The functional integral over the external field in Eq.~\ref{eq:noise_dist} 
then reduces to a standard integral over a single Gaussian distribution
with a fixed variance of the annealed disorder, $\sigma_{\rm ad}^2$.
The final result of this procedure is that the reduced density matrix is
given by 
$\hat \rho(\beta) = \langle e^{-\beta \hat H}\rangle/{\langle Z \rangle}$,
where the angular brackets denote averaging with respect to the resulting 
Gaussian distribution of realizations of the slow bath.
As with the case of general noise discussed above, the averaging here arises 
from tracing out the bath degrees of freedom.
Contrast this situation with that of of quenched disorder in which 
one is led to the alternate expression for the reduced density matrix, 
$\hat \rho(\beta) = \langle e^{-\beta \hat H}/Z \rangle$.
The averaging here is carried out over the distribution of Hamiltonians.
It is clear that these two approaches are not equivalent.
They lead to qualitatively different behavior at low temperatures.

\begin{figure}
    \includegraphics*[width=0.85\textwidth]{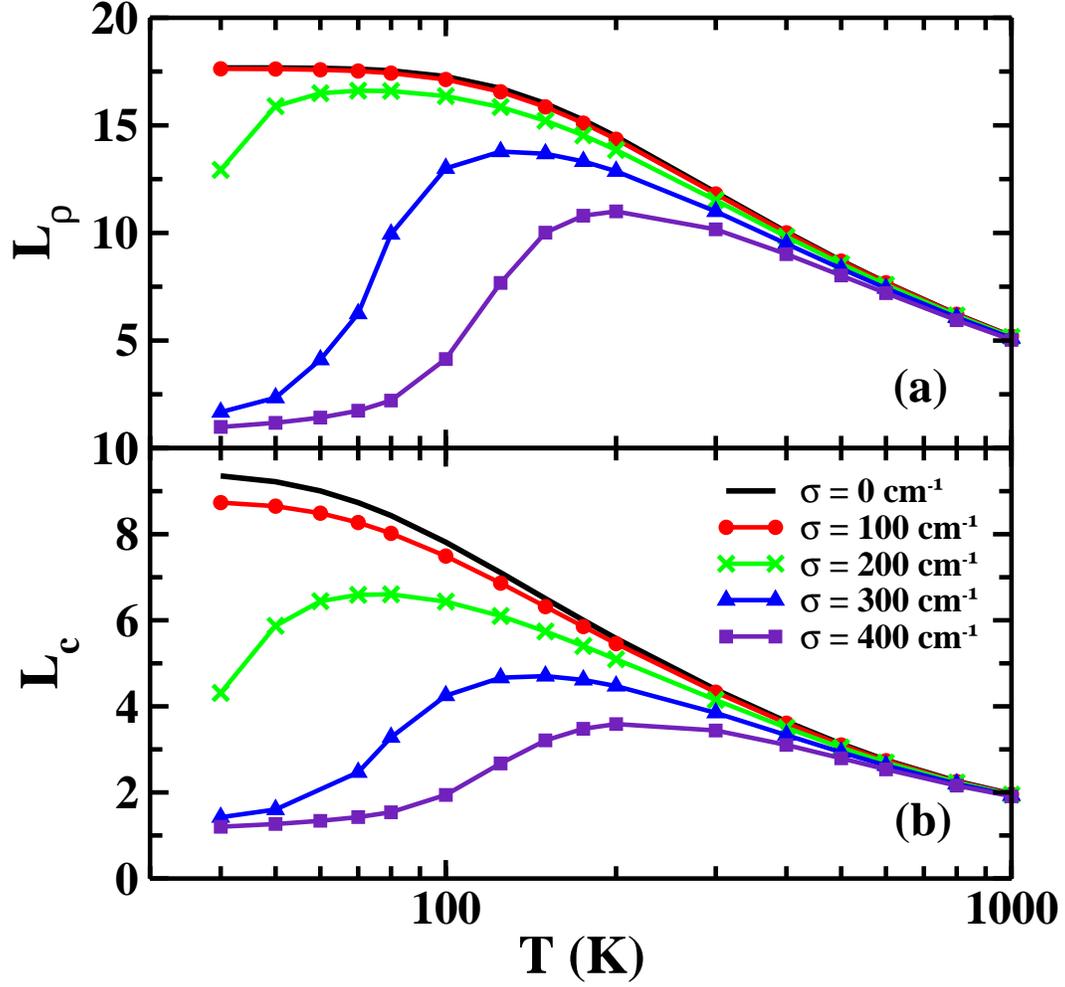}
    \caption{(color online)
    The coherence lengths $L_{\rho}$ (a), 
    and $L_c$ (b), calculated for LH2 as a function 
    of temperature for increasing strength of the annealed disorder.
    The solid (black) line is obtained from the Boltzmann populations 
    without any disorder.
    The widths of the annealed disorder distribution of
    $100$, $200$, $300$, and $400$ cm$^{-1}$
    are indicated by the (red) line with dots, (green) line with crosses, 
    (blue) line with triangles, and (purple) line with squares, respectively.
   }
   \label{fig:LH2_ab}
\end{figure}

The adiabatic bath model has been extensively analyzed in the context of
the two-level system.\cite{chandler91}
In this model, there exists a critical value of the system parameters 
at which self trapping occurs.
The system becomes localized when $\beta \sigma_{\rm ad}^2/J > 1$, 
where $J$ is the coupling between the two levels.\cite{chandler91}
This transition persists for larger systems and is shown in 
Fig.~\ref{fig:LH2_ab} for LH2.
Provided that the strength of the annealed disorder is greater than 
a critical value, there is a transition to a localized state at low temperature
as indicated by the dramatic decrease seen in the coherence length measures.
In contrast, the case of quenched disorder shown previously in 
Fig.~\ref{fig:LH2_coherences} displayed the opposite behavior.
There, both of the coherence length measures increased monotonically 
with decreasing temperature.
It should be noted that a proper treatment of the limit of $K(\tau)$ 
for small $\omega_{c}$ leads a variance of the adiabatic bath that 
depends on both the temperature and reorganization energy.
However, in order to be consistent with the treatment of static disorder
and the analysis of Ref.~\onlinecite{chandler91}, here we simply 
take $\sigma_{\rm ad}$ to be constant.

It is unlikely, however, that the adiabatic bath plays a large role 
in light harvesting systems as the migration of the exciton 
from one independent complex to another occurs on relatively fast time scales.
Observing the effects of annealed disorder may be possible in other cases such
as in extended J-aggregates where exciton lifetimes and diffusion lengths
can be much longer than in light harvesting systems.
It is known that the coherence length is related to both the superradiance
and the Stokes shift at low temperatures.
Observing transitions in these or other related quantities could be an 
indication that the slow bath fluctuations are non-negligible.
Regardless of the implications, the static disorder which is most often 
referred to in discussions of light harvesting systems should not 
be described in terms of slow bath fluctuations.

\subsection{FMO}

\begin{figure}
   \includegraphics*[width=0.85\textwidth]{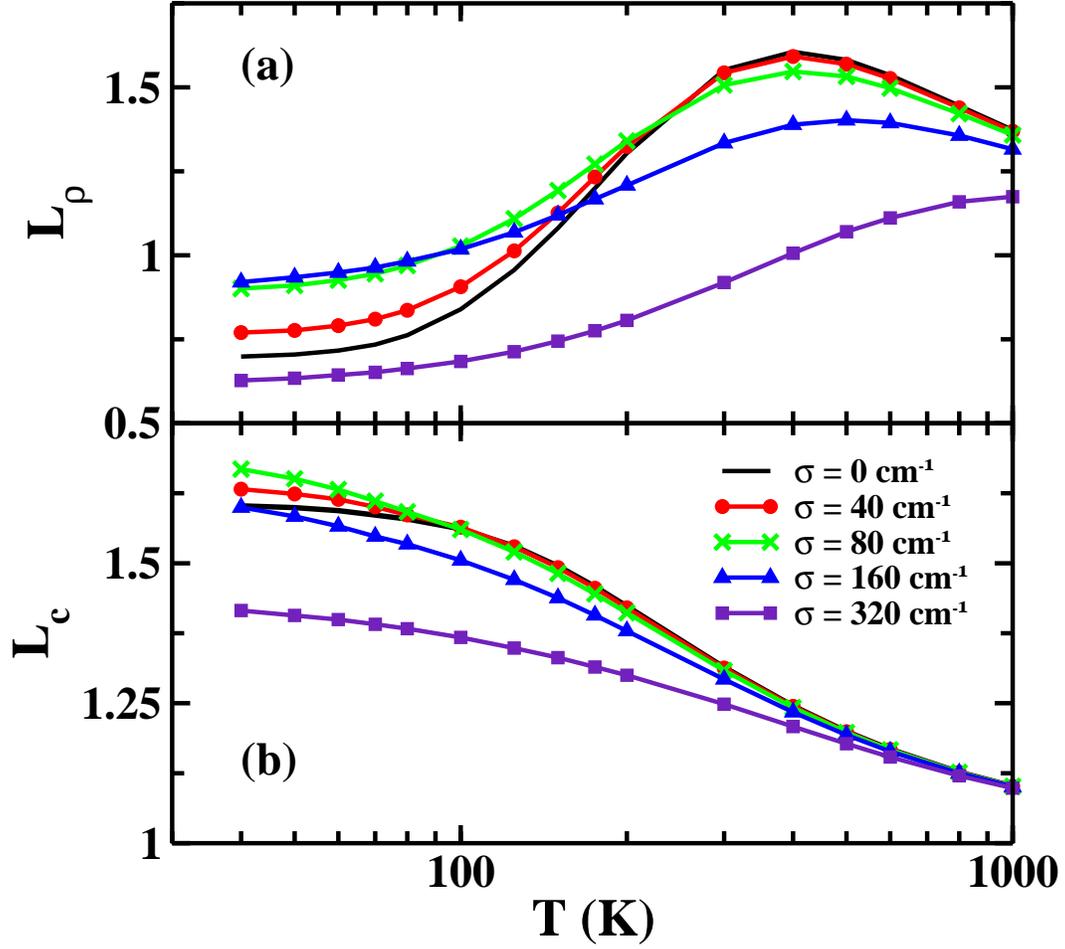}
   \caption{(color online)
    The coherence lengths $L_{\rho}$ (a) and $L_c$ (b), 
    calculated for FMO as a
    function of temperature for increasing static disorder.
    The solid (black) line is obtained from the Boltzmann populations 
    without any static disorder.
    The widths of the static disorder distribution of
    $40$, $80$, $160$, and $320$ cm$^{-1}$
    are indicated by the (red) line with dots, (green) line with crosses, 
    (blue) line with triangles, and (purple) line with squares, respectively.
   }
   \label{fig:FMO_coherences}
\end{figure}

Calculations of the coherence lengths performed for the FMO complex with 
static disorder are shown in Fig.~\ref{fig:FMO_coherences}.
As with LH2, estimates for the width of the static disorder distribution
vary widely although the value of $80$ cm$^{-1}$ has been used
to fit several experimental results.\cite{milder10}
The coherences lengths in FMO display behavior that is qualitatively 
different from those seen in LH2 in Fig.~\ref{fig:LH2_coherences}.
Even without static disorder, neither of the coherence lengths  
continually increase to their respective maximal value as the temperature 
is lowered as was observed for LH2. 
Similar to the results presented for the biased two level system in 
Fig.~\ref{fig:TLS_IPR}, $L_{\rho}$ reaches a maximum value and then decreases 
as a function of the temperature.
Additionally, at low temperatures it is seen that the static disorder 
can lead to an increase in the coherence length as was also observed in
the biased two level system.
As discussed previously, the peak in the $L_{\rho}$ and values less than 
one are an indication of the onset of localization at the lowest energy site.
Direct inspection of the reduced density matrix confirms that this is the case.
Regardless of the definition, both estimates yield rather small values 
for the coherence length that extends over one or two chromophores.
As mentioned previously, FMO functions as an energy funnel in 
photosynthetic systems which efficiently transports energy from 
the antenna complex to the reaction center.
Therefore, the equilibrium state will rarely be reached in this system.
Nevertheless, FMO is considered here as simply a reasonable model 
for general disordered excitonic systems.

\begin{figure}
   \includegraphics*[width=0.85\textwidth]{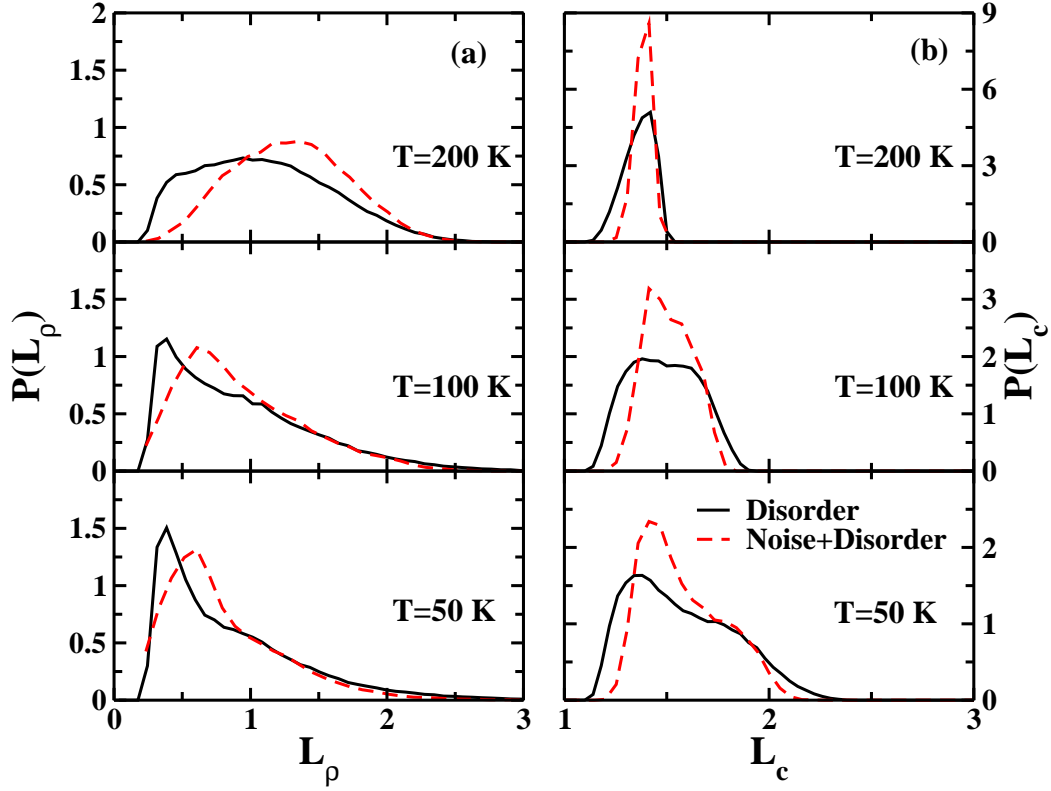}
   \caption{(color online)
    The distribution of the the coherence lengths, $L_{\rho}$ (a), 
    and $L_c$ (b) for FMO. 
    The top, middle and bottom panels correspond to temperatures of 
    $200$, $100$ and $50$ K.
    The solid (black) lines represent the distributions obtained with 
    static disorder only with a width of $80$ cm$^{-1}$.
    The dashed (red) lines display the results of disorder and noise
    with a reorganization energy of $35$ cm$^{-1}$.
   }
   \label{fig:FMO_hist}
\end{figure}

The coherence length distributions of FMO calculated with both disorder
and noise are shown in Fig.~\ref{fig:FMO_hist}.
As with the case of LH2, the temperature effect is seen to have 
the largest impact on the coherence length distributions.
It has been noted above that the presence of disorder can increase
the coherence length in FMO.
Fig.~\ref{fig:FMO_hist} demonstrates that the addition of noise
may increase the coherence length even further.
In all cases, the distribution of noise and disorder compared with that 
of disorder alone is shifted to higher values of the coherence length.
The origin of the maximum that is observed in Fig.~\ref{fig:FMO_coherences}(a)
can be seen from the differences between the distributions of $L_{\rho}$ at
$200$ K and $100$ K.
At $100$ K, a significant portion of the distribution of coherence lengths
is composed of values of $L_{\rho}$ less than $1$ which is indicative of 
localization on a single site.

\section{Concluding remarks}

A general and efficient method for computing the exact equilibrium matrix 
for systems governed by a system-bath Hamiltonian was presented.
This approach has several advantages over standard implementations of 
the imaginary time path integral approach.
First, due to a Hubbard-Stratonovich transformation,
the imaginary time convolution that appears in the influence functional 
is replaced by an additional functional integral over an auxiliary Gaussian
field.
The latter is readily amenable to a straightforward importance sampling
Monte Carlo procedure.
Additionally this allows one to compute the entire reduced density matrix 
from a single simulation, whereas in standard path integral treatments 
each element of the density matrix must be evaluated separately.
As was demonstrated above, the simplicity and versatility of this 
approach allows for the treatment of both quenched and static disorder,
as well as general thermal noise all within the same framework 
outlined in Sec.~\ref{sec:Methods}.

Applying this technique, we presented exact results for the equilibrium 
reduced density matrix and the coherence lengths in FMO and LH2.
For typical values of the system-bath coupling in these systems, 
it was demonstrated that the system and bath are entangled in the
single exciton manifold, even at room temperature.
As seen from Figs.~\ref{fig:populations} and~\ref{fig:rho1i}, 
the exciton populations are only approximately given by a Boltzmann
distribution of the bare system, and this approximation becomes progressively
worse as the system-bath coupling increases and the temperature is lowered.

The role of static disorder, thermal noise and temperature 
on the coherence length was then systematically investigated 
for two commonly used measures of this quantity.
Comparing Fig.~\ref{fig:TLS_IPR} with Figs.~\ref{fig:LH2_coherences} 
and~\ref{fig:FMO_coherences} it is seen that the qualitative influence 
of static disorder on LH2 and FMO is well described by that of the 
respective symmetric and biased two level system.
An extensive analysis for the latter was presented in Sec.~\ref{sec:TLS}.
For symmetric systems, disorder only leads to a decrease in the coherence
length, whereas in biased systems, disorder may increase the coherence length. 
There are some important differences between the role noise and disorder, 
particularly at low temperatures.
For example, static disorder leads to a plateau in the coherence length 
that becomes increasingly pronounced with increasing strength of the disorder
at low temperatures.
The addition of thermal noise, however, can prevent this feature from 
occurring.\cite{kamishima07}
The combined effect of both noise and disorder shown in Figs.~\ref{fig:LH2_hist}
and~\ref{fig:FMO_hist} is different from either of them alone.
In unbiased systems, increasing the temperature narrows the distribution 
of the coherence lengths,
whereas static disorder leads to both broadening and skewing.
The presence of noise results in an additional shift of the distribution 
towards the localization.
However, for biased systems such as FMO, disorder can increase the 
coherence length, 
and the additional presence of noise may increase it further.
Finally the influence of the environment as modeled either 
by static disorder or a slow bath was compared in 
Figs.~\ref{fig:LH2_coherences} and~\ref{fig:LH2_ab}.
These two scenarios are not equivalent as is often claimed,
and they lead to qualitatively different behavior at low temperature.
Annealed disorder (a slow bath) leads to a sharp transition to localization at 
low temperature whereas quenched disorder does not.

While the two measures of the coherence length in 
Eqns.~\ref{eq:IPR} and~\ref{eq:L} provide the 
same qualitative picture of the effect of the environment on the coherence
length at high temperatures, some significant differences appear
in the low temperature regime.
For biased systems such as FMO, $L_{\rho}$ shows a maximum in the coherence
length as a function of temperature whereas $L_c$ does not.
In symmetric systems, $L_{\rho}$ leads to a plateau in the 
coherence length at low temperatures that is much more pronounced than
that of $L_c$.
Additionally, $L_{\rho}$ predicts that LH2 is almost completely coherent
at low temperatures regardless of the strength of the thermal noise.
These qualitative differences should serve as a note of caution when
using such coarse measures to characterize the coherence length in excitonic
systems.

Due to the generality of the numerical scheme presented here, 
there are many possible extensions to this work.
The application of this path integral technique to the study of the 
coherence length in larger systems such as J-aggregates, the chlorosome,
and nanotubes is currently being investigated.
In a forthcoming publication, the path-integral results provide a
benchmark for various approximate methods including the polaron transformation 
and its variational form.
Additionally, this approach can provide exact results for the 
thermal entanglement at finite temperatures in quantum information 
and quantum computing applications.
Finally we mention the possibility of analytically continuing imaginary
time correlation functions in order to obtain the corresponding real time 
quantities.
For example, this would allow for the exact calculation of diffusion 
coefficients in rather large systems.
These topics will be the focus of future publications.

\section{Acknowledgments}

We thank Johan Str{\"u}mpfer for providing the crystal structure of
LH2 needed for constructing the Hamiltonian used in this work.
This work was supported by grants from the National Science Foundation
(grant number CHE-1112825),
DARPA, 
and the Center for Excitonics at MIT funded by the Department of Energy
(grant number DE-SC0001088).
Additionally, support from the Singapore National Research Foundation 
through the Competitive Research Programme (CRP) under Project 
No.~NRF-CRP5-2009-04 is gratefully acknowledged.
J Cao is supported as part of the Center of Excitonics, an Energy Frontier
Research Center funded by the US Department of Energy, Officie of Science,
Office of Basic Energy Sciences under Award No. DE-SC0001088.

\appendix

\section{Discrete systems}\label{sec:Discrete_PI}

The procedure leading to Eq.~\ref{eq:Bloch} in Sec.~\ref{sec:Methods}
is most easily generalized to discrete systems through the mapping
formalism.\cite{stock97,thoss99}
In this approach, the discrete levels of the system Hamiltonian are mapped 
to continuous bosonic creation and annihilation variables through the 
relations,
$\left|n\right\rangle\left\langle m\right| \rightarrow a_n^\dagger a_m$. 
The path integrals over the system and bath can then be constructed 
in a mixed representation using the usual coordinate states for 
the bath degrees of freedom
and the coherent state representation for the bosonic modes.
This approach has been successfully applied, for example, to model the dynamics
of a two level system coupled to a dissipative vibrational degree of 
freedom.\cite{novikov04}
The construction of the system action for the bosonic modes requires
some care but is extensively discussed in 
Refs.~\onlinecite{kleinert04} and~\onlinecite{su00}. 
However, as with the case of the continuous Hamiltonians discussed in the main
text, the explicit form of the system action is never required.
With these preliminaries, the procedure outlined in Sec.~\ref{sec:Methods}
proceeds identically.
The influence functional follows the continuous form before
except with the system coordinate $f(q)$ replaced by the proper 
coupling (generally it is one of the sites $a_n^\dagger a_n$),
and the Hubbard-Stratonovich transformation in unaffected.
At the end of the calculation one can map the bosonic variables
back to the original discrete levels, and 
finally obtain the result of Eq.~\ref{eq:Bloch}.

\section{FMO Hamiltonian}\label{sec:FMO_ham}

The Hamiltonian for FMO is taken from Ref.~\onlinecite{cho05}.
The specific values (in cm$^{-1}$) are given by 
\begin{equation}
   H_{\rm FMO} =
\begin{pmatrix}
 280 & -106 &   8 &  -5 &   6 &  -8 &  -4 \\ 
-106 &  420 &  28 &   6 &   2 &  13 &   1 \\
   8 &   28 &   0 & -62 &  -1 &  -9 &  17 \\
  -5 &    6 & -62 & 175 & -70 & -19 & -57 \\
   6 &    2 &  -1 & -70 & 320 &  40 &  -2 \\
  -8 &   13 &  -9 & -19 &  40 & 360 &  32 \\
  -4 &    1 &  17 & -57 &  -2 &  32 & 260
\end{pmatrix}\;.
\end{equation}

\bibliography{j,halcyon,centroid,liquid,light_harvesting,sc,diffusion}

\newpage

\begin{figure}
   \includegraphics*[width=0.85\textwidth]{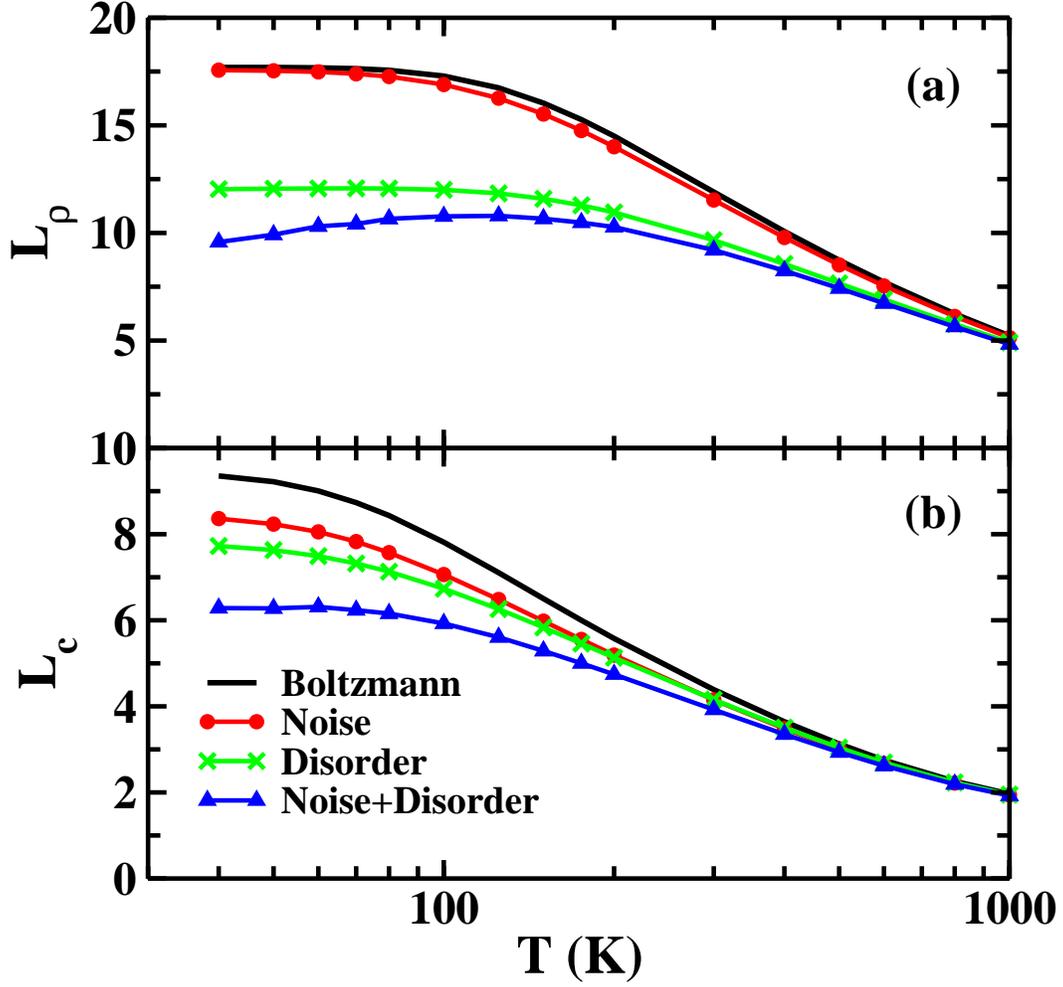}
   \caption{(color online)
    The coherence lengths, $L_{\rho}$ (a), and $L_c$ (b), 
    calculated for LH2 as a function of temperature.
    The solid (black) line is obtained 
    from the Boltzmann populations without noise or disorder.
    The (red) line with dots is obtained from calculations with noise only
    using the spectral density of Eq.~\ref{eq:sd_olbrich}, 
    the (green) line with crosses is obtained from calculations with 
    static disorder only with a width of $200$ cm$^{-1}$, 
    and the (blue) line with triangles is from calculations with both 
    noise and disorder.
   }
   \label{fig:LH2_noise_disorder}
\end{figure}

\end{document}